\documentclass[%
reprint,
superscriptaddress,
amsmath,amssymb,
aps,
pra,
longbibliography
]{revtex4-1}

\usepackage{graphicx}
\usepackage{dcolumn}
\usepackage{bm}

\usepackage{amsmath,amssymb}
\usepackage{graphicx}
 \usepackage{color}
\usepackage[binary-units=true]{siunitx}

\newcommand{\ii}{\textrm{i}}
\newcommand{\ee}{\textrm{e}}

\newcommand{\dd}{\textrm{d}}
\newcommand{\bigO}{\textrm{O}}
\newcommand{\inc}{\textrm{in}}

\newcommand{\grad}{\textrm{grad}}

\DeclareMathOperator{\re}{Re}
\DeclareMathOperator{\im}{Im}

\renewcommand{\vec}[1]{\bm{#1}}

\begin{document}


\title{Nonlinear interaction of acoustic waves with a spheroidal particle: radiation force and torque effects}



\author{Everton B. Lima}%
\affiliation{%
	Physical Acoustics Group,
	Instituto de F\'isica,
	Universidade Federal de Alagoas, 
	Macei\'o, AL 57072-970, Brazil}%
\author{Jos\'e P. Le\~ao-Neto}
\affiliation{Campus Arapiraca/Unidade de Ensino Penedo, Universidade Federal de Alagoas, Penedo, Alagoas
	57200-000, Brazil}
\author{Alisson S. Marques}
\affiliation{%
	Physical Acoustics Group,
	Instituto de F\'isica,
	Universidade Federal de Alagoas, 
	Macei\'o, AL 57072-970, Brazil}
\author{Gicl\^enio C. Silva}
\affiliation{%
	Physical Acoustics Group,
	Instituto de F\'isica,
	Universidade Federal de Alagoas, 
	Macei\'o, AL 57072-970, Brazil}%
\author{Jos\'e H. Lopes}
\affiliation{Grupo de F\'isica da Mat\'eria Condensada, N\'ucleo de Ci\^encias Exatas,
	Universidade Federal de Alagoas, Arapiraca, AL 57309-005, Brazil}
\author{Glauber T. Silva}\email{gtomaz@fis.ufal.br}
\affiliation{%
	Physical Acoustics Group,
	Instituto de F\'isica,
	Universidade Federal de Alagoas, 
	Macei\'o, AL 57072-970, Brazil}%


\date{\today}

\begin{abstract}
The nonlinear interaction of a time-harmonic acoustic wave with an anisotropic particle gives rise to the radiation force and torque effects.
These phenomena are at the heart of the acoustofluidics technology, where 
microparticles such as cells and microorganisms are acoustically manipulated. 
We present a theoretical model considering a generic acoustic beam interacting with a subwavelength spheroidal particle in a nonviscous fluid.
Concise analytical expressions of the radiation force and torque are obtained in the scattering dipole approximation.
The radiation force is given in terms of a gradient and scattering force; 
while the radiation torque has two fundamental contributions, namely, the momentum arm and
acoustic spin (spin-torque effect).
As a practical example, we use the theory to describe the interaction of two crossed plane waves and a prolate spheroidal particle.
The results reveal the particle is transversely trapped in a pressure node and is axially pushed by the radiation force.
Also, the momentum arm aligns the particle in the axial direction.
At certain specific positions, only the spin-torque occurs.
Our findings are remarkably consistent with finite-element simulations.
The success of our model enables its use as an investigation tool for the manipulation of anisotropic microparticles in acoustofluidics.
\end{abstract}

\maketitle


\section{Introduction}
The behavior of microparticles under an ultrasonic acoustic wave has been extensively analyzed in micro-acoustofluidic devices~\cite{Ozcelik2018,Wu2019}.
Notable examples include separation of circulating tumor cells~\cite{Li2015},
cell and microparticle patterning~\cite{Collins2015,Silva2019},
assess the membrane elasticity of cell by acoustic deformation~\cite{Mishra2014,Silva2019a}, and selective acoustic tweezer~\cite{Baudoin2019}.
The nonlinear wave-particle interaction gives rise to
the acoustic radiation forces and torques phenomena~\cite{Baudoin2019b}.
The careful control of these effects enables particle handling in micrometer-sized cavities or microchannels
with a plethora of applications in biotechnology and analytical chemistry.

The acoustic radiation forces and torques are commonly
investigated considering isotropic particles, i.e., with a spherical shape.
In reality, the morphology of most cells and other microorganisms have a degree of asymmetry.
Prominent examples of acoustofluidic systems for manipulation of asymmetric particles include glass fibers~\cite{Schwarz2012}, \emph{Escherichia coli} bacterium~\cite{Ai2013}, red blood cells~\cite{Jakobsson2014}, microfibers~\cite{Schwarz2015}, and alumina microdisks~\cite{Garbin2015}.
Other experiments have been performed in acoustic levitation systems in air~\cite{Foresti2013,Marzo2015,Hong2017}.
Understanding how acoustic forces and torques develop on anisotropic microparticles is key to dynamic analysis, as well as to devise new applications of acoustofluidic methods.
Additionally, these phenomena seem to have a crucial role in the propelling mechanisms of microswimmers under an ultrasound field~\cite{Wang2012,Ren2019}.

At first glance, the available alternative to model the wave interaction with anisotropic particles is the use of numerical techniques, such as the finite~\cite{Schwarz2012,Glynne-Jones2013,Garbin2015,Hahn2015,Greve2018} and boundary~\cite{Wijaya2015} element methods, 
Born approximation~\cite{Jerome2019},
numerical quadrature~\cite{Mitri2015,Mitri2016a},
and $T$-matrix approach~\cite{Gong2019a}.
In general, numerical methods demand high-performance computing and high memory usage for three-dimensional simulations.
Moreover, 
it is also challenging to determine the behavior of the wave-particle system as one or more parameters vary continuously.

Early studies involving anisotropic particles dealt with the radiation torque problem on circular disks~\cite{Kotani1933,King1935,Keller1957,Maidanik1958}.
Some other investigations have been surveyed in Ref.~\cite{Marston2016}. 
More recently, efforts have  been devoted to describing the acoustic radiation force~\cite{Marston2006a,Silva2018} and 
torque~\cite{Fan2008,Leao-Neto2020,Lopes2020} on spheroidal particles.
These analyses rely on the partial-wave expansion of the acoustic fields.
In this method,
the expansion coefficients of the incoming beam (beam-shape coefficients) should be known \emph{a priori}~\cite{Baresch2013,Mitri2014,Silva2015a,Gong2017,Leao-Neto2017,Zhang2018a}.
Numerical schemes can also be employed to compute the beam-shape coefficients~\cite{Silva2011a,Mitri2011,Silva2013,Lopes2016,Gong2019}, and even experimental methods~\cite{Zhao2019}.
Regardless these studies,
the ultrasonic waves produced in acoustofluidic devices have a complex spatial form (structured waves), and
the corresponding beam-shape coefficients are generally unknown.
In the case of isotropic particles, the radiation force caused by structured waves can be easily computed using the Gorkov's theory~\cite{Gorkov1962}, which requires only the incident pressure and fluid velocity, either analytically or numerically.
A similar approach was developed for the radiation torque on a spherical particle~\cite{Silva2014}.

The purpose of this article is to model the radiation force and torque imparted on a subwavelength spheroid by an arbitrary acoustic beam.
Our approach is based on the general expressions of the radiation force~\cite{Silva2018} and torque~\cite{Lopes2020,Leao-Neto2020} that depend on the expansion coefficients of the incident and scattering waves., e.g., the beam-shape and scattering coefficients.
These expressions are exact to the dipole moment for the scattering wave.
The scattering coefficients are obtained through the boundary conditions on the particle surface~\cite{Silva2018}.
Also, the relationship between the beam-shape coefficients and the incident acoustic fields (pressure and fluid velocity) is established.
Strikingly, the final expressions of the radiation force and torque are given in terms of the incoming fields evaluated at the geometric center of the particle.
Despite being developed for a rigid spheroid, the theory can be readily adapted to accommodate the elasticity and absorption of the particle, as well as the surrounding fluid viscosity. 
These effects have already been investigated for isotropic particles~\cite{Leao-Neto2016,Zhang2014}.

Our model is used for the analysis of a spheroidal particle interacting with two crossed plane waves at right angle.
This structured beam forms a transverse standing wave and an axial (perpendicular) traveling wave.
The model predicts the radiation force pushes the particle to a transverse pressure node or antinode. 
The radiation torque aligns the particle at a node in broadside orientation. 
In contrast, the orientation in an antinode is along the axis of the standing wave.
Additionally, the axial radiation force pushes the particle in the direction parallel to the pressure node.
Lastly, the model predictions are verified against finite-element (FE) simulations.
We find an excellent agreement between the theoretical and numerical results.

\section{Theoretical model}

\subsection{Governing equations}
Consider a fluid of infinite extension characterized by an ambient density $\rho_0$, adiabatic speed of sound $c_0$, and compressibility $\beta_0=1/\rho_0 c_0^2$.
Our analysis is restricted to acoustic fields of time 
harmonic dependence $\ee^{-\ii \omega t}$, with $\omega=2\pi f$ and $f$ being the angular and linear frequencies, respectively.
The corresponding  wavenumber is $k=\omega/c_0=2\pi/\lambda$, where  $\lambda$ is the acoustic wavelength.
The acoustic pressure and fluid velocity are expressed using
the complex-phase representation,  $p(\vec{r},t)=p(\vec{r})\ee^{-\ii \omega t}$ and 
$\vec{v}(\vec{r},t)=\vec{v}(\vec{r})\ee^{-\ii \omega t}$,
respectively.
In Cartesian coordinates, the position vector and fluid velocity are
$r_i \vec{e}_i$ (with $i=x,y,z$) and
$\vec{v}=v_i\vec{e}_i$,
where the $\vec{e}_i$ is the Cartesian unit vector.
The summation over repeated indexes is
automatically assumed hereafter.
We also adopt the notation $(r_x,r_y,r_z)=(x,y,z)$.

In the inviscid limit, the wave dynamics is modeled by the well-known linear acoustic equations
\begin{subequations}
	\label{dynamic_eqs}
	\begin{align}
	\label{euler}
	\vec{v} &=  -\frac{\ii}{\rho_0c_0 k}\nabla p,\\
	\left(\nabla^2 + k^2\right) p &= 0,
	\label{helm_pressure}\\
	\label{irrot}
	\nabla \times \vec{v}& = \vec{0}.
	\end{align}
\end{subequations}
The time-dependent term $\ee^{-\ii \omega t}$ is omitted for simplicity.
Equation~\eqref{irrot} is obtained by taking the rotational of Eq.~\eqref{euler}.

In the presence of an inclusion, such as a 
particle, an incident wave is scattered by the inclusion. 
The boundary condition for the scattered pressure $p_\text{sc}$ at the farfield is the Sommerfeld radiation condition.
In spherical coordinates $(r,\theta,\varphi)$, this means
\begin{equation}
\label{Rcondition}
\lim_{r\rightarrow\infty} r\left(\partial_r - \ii k\right) p_\text{sc} = 0.
\end{equation}
The radiation condition singles out only the solution which represents ``outgoing'' waves. 

Another boundary condition comes from
considering the particle as a rigid body.
In this case,
the normal component of the total velocity of the fluid, i.e. incident plus scattering contributions, should vanish
on the particle surface $S_0$,
\begin{equation}
\vec{n}\cdot \vec{v}|_{\vec{r}\in S_0} = 0.
\label{velBoundary}
\end{equation}
Here $\vec{n}$ is the outward normal unit-vector on $S_0$.
\begin{figure}
	\includegraphics[scale=.94]{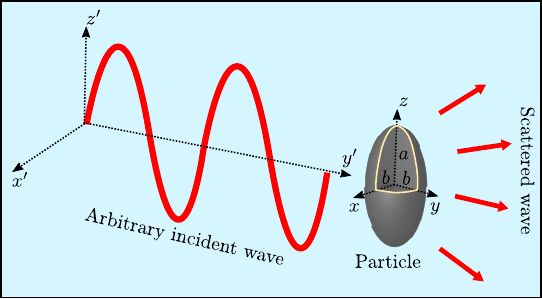}
	\caption{
		The acoustic wave scattering by the spheroidal particle in a fluid medium. The particle has a major and minor semiaxis denoted by $a$ and $b$. The arbitrary incoming beam is depicted as a red sinusoidal line. The scattered waves are represented by red arrows. 
		The geometric center of the spheroid defines the particle frame of reference, $\vec{r}=(x,y,z)$.
		The laboratory frame of reference is the primed Cartesian axes.
		\label{fig:problem}}
\end{figure}

\subsection{Prolate spheroidal particle}
A prolate spheroidal particle 
with a major and minor axis denoted by $2a$ and $2b$,
is placed in the wave path of an arbitrary incoming beam as depicted in Fig.~\ref{fig:problem}.
Its geometric center defines the particle coordinate system 
$(x,y,z)$.
The particle surface is conveniently described in prolate spheroidal coordinates whose connection 
the Cartesian system is expressed by
\begin{subequations}
	\label{CtoS}
	\begin{align}
	x &=\frac{d}{2}\sqrt{(\xi^2-1)(1-\eta^2)}\cos \varphi,\\
	y &=\frac{d}{2}\sqrt{(\xi^2-1)(1-\eta^2)}\sin \varphi,\\
	z &=\frac{d\xi \eta}{2},
	\end{align}
\end{subequations}
where 
$\xi\geq 1$ is the spheroidal radial coordinate,
$-1\leq\eta\leq1$, and  $0\leq\varphi\leq2\pi$ is azimuth angle, with the interfocal distance being
\begin{equation}
\label{interf}
d=2\sqrt{a^2-b^2}.
\end{equation}
The particle surface is given by
\begin{equation}
\label{AS}
\xi=\xi_0=\left[1-\left(\frac{b}{a}\right)^{2}\right]^{-1/2},
\end{equation}
where $\xi_0$ is regarded as the  geometric parameter of the particle.
The particle orientation follows the $z$-direction, $\vec{d} = d \, \vec{e}_{z}$. 
Note that we recover a sphere of radius $r_0$ by setting
\begin{equation}
\label{sphere}
d\rightarrow 0, \quad \xi_0\rightarrow \infty, \quad \frac{\xi_0 d}{2}\rightarrow r_0.
\end{equation}

The particle dynamics is better analyzed in a fixed laboratory system
$(x',y',z')$.
In this reference frame, the particle orientation angle $\alpha$ is expressed by $\vec{d}\cdot \vec{e}_{z'} = d \cos\alpha$.

\subsection{Acoustic scattering}
The incoming and scattered pressure can be expressed by the  partial wave expansion in spherical coordinates $(r,\theta,\varphi)$,
\begin{subequations}
	\label{phis}
\begin{align}
\label{phi_in}
    p_\inc &=p_0
	\sum_{n=0}^\infty \sum_{m=-n}^n {a}_{nm}
	j_n(kr)Y_n^m(\theta,\varphi),\\
	p_\text{sc}  &=p_0
	\sum_{n=0}^\infty \sum_{m=-n}^n {a}_{nm}s_{nm}
	h_n(kr)Y_n^m(\theta,\varphi),
	\label{psc}
\end{align}
\end{subequations}
in which $p_0$ is the peak pressure,  $j_n$ is the $n$th-order spherical Bessel function, 
$h_n$ is the spherical Hankel function of the first-type,
and
$Y_n^m$ is the spherical harmonics of $n$th-order and $m$th-degree.
The quantity $a_{nm}$ is known as the beam-shape coefficient. 
The scattering coefficients $s_{nm}$ can be obtained from the boundary condition in Eq.~\eqref{velBoundary}.

As we advance to consider a particle much smaller than the wavelength, i.e., the so-called long-wavelength limit,
only the monopole ($n=0$) and dipole ($n=1$)
modes of the multipole expansion in \eqref{psc}
suffice to describe the acoustic scattering~\cite{Pierce2019}.
We define the smallness parameter of the particle as 
\begin{equation}
\label{kd}
\epsilon= \frac{\xi_0  k d}{2}   = k a \ll 1.
\end{equation} 
We shall use $\epsilon$ as the expansion parameter in the long-wavelength approximation.

Using the partial wave expansion in spheroidal coordinates $(\xi,\eta,\varphi)$, one can show that the monopole and dipole scattering coefficients of a rigid spheroid are given by~\cite{Silva2018}
\begin{subequations}
	\label{scatt_spheroid}
	\begin{align}
	s_{00} &= -\frac{\ii {\epsilon}^3}{3} f_{00}
	- \frac{{\epsilon}^6}{9} f_{00}^2,\\
	s_{10} &=\frac{\ii{\epsilon}^3}{6}f_{10}-\frac{{\epsilon}^6}{36}f_{10}^2 ,\\
	s_{1,-1}
	&=s_{11}=\frac{\ii{\epsilon}^3}{12}f_{11}-\frac{{\epsilon}^6}{144}f_{11}^2,
	\end{align}
\end{subequations}
with $\ii$ being the imaginary unit.
The dipole coefficients of the modes perpendicular to the axial direction ($s_{1,-1}$ and $s_{11}$) 
are degenerated due to the axial symmetry of the particle.
The scattering factors $f_{00}$, $f_{10}$, and $f_{11}$ of Ref.~\cite{Silva2018} are re-written here as
\begin{subequations}
	\label{factors}
	\begin{align}
	f_{00} &=   1 - \xi_0^{-2},\\
	f_{10} &=\frac{2}{3\xi_0^3}\left[\frac{\xi_0}{\xi_0^2-1}-\ln\left(\frac{\xi_0+1}{\sqrt{\xi_0^2-1}}\right)\right]^{-1} ,\\
	f_{11}
	&=\frac{8}{3 \xi_0^3}\left[\frac{2-\xi_0^2}{\xi_0(\xi_0^2-1)}+\ln\left(\frac{\xi_0+1}{\sqrt{\xi_0^2-1}}\right)\right]^{-1}.
	\end{align}
\end{subequations}
These functions depend solely on the geometric factor $\xi_0$.

To recover the scattering coefficients of a  spherical particle, we take the limit $\xi_0\rightarrow \infty$ in
\eqref{factors}, which yields 
\begin{equation}
\label{f_sphere}
f_{00} = 1, \quad f_{10}=1, \quad f_{11}=2.
\end{equation}
It immediately follows from \eqref{scatt_spheroid} that
the dipole scattering coefficients of a sphere are all degenerated,  as expected,
\begin{equation}
\label{scatt_sphere}
s_{10} = s_{1,-1} = s_{11}, \quad\text{(sphere)}.
\end{equation}

\subsection{Multipole expansion of the incident beam}
The beam-shape coefficients of an incident beam is obtained
as follows. 
Multiplying Eq.~\eqref{phi_in} by $Y_n^{m*}$, integrating the result
over the unit-sphere, and using the orthogonal relation of the spherical harmonics
$\int_0^\pi\int_0^{2\pi} Y_n^m(\theta,\varphi)Y_{n'}^{{m'}*}(\theta,\varphi)\sin\theta\,\dd \theta\,\dd \varphi =  \delta_{n{n'}}\delta_{m{m'}}$ (with $\delta_{nm}$ being the Kronecker delta function), we find~\cite{Silva2011a}
\begin{align}
\nonumber
    a_{nm} &=
    \\
    &\frac{1}{p_0\, j_n(k r)}\int_0^\pi\int_0^{2\pi} p_\text{in}(k r, \theta, \varphi)
    Y_n^{m*}(\theta,\varphi)\sin\theta\, \dd \theta\, \dd \varphi,
    \label{anm2}
\end{align}
where asterisk denotes complex conjugation.
The beam-shape coefficient can be calculated for any radial distance $kr$.
In particular, one can relate them with the acoustic pressure evaluated in the origin of the particle frame, $\vec{r}=\vec{0}$.
We anticipate here that the acoustic radiation force depends
on the beam-shape coefficients to the quadrupole approximation~\cite{Silva2014}. 
Therefore, we truncate the Taylor expansion of the incident pressure around $\vec{r}=\vec{0}$ as
\begin{equation}
    p_\text{in}(\vec{r}) 
    = p_\text{in}(\vec{0})
    + r_i \partial_i p_\text{in}|_{\vec{r}=\vec{0}}+ \frac{1}{2!}r_ir_j \partial_i\partial_j
    p_\text{in}|_{\vec{r}=\vec{0}},
    \label{phiT}
\end{equation}
with $i,j=x,y,z$.
Before proceeding, 
we  need  the asymptotic expansion of the spherical Bessel function for $r\rightarrow0$,
\begin{equation}
  j_n(k r) \approx  
  \frac{\sqrt{\pi}}{2^{n+1} \Gamma(n+3/2) }
 (k r)^n,
 \label{jT}
\end{equation}
in which $\Gamma(n)$ is the gamma function.
Replacing Eqs.~\eqref{phiT}
and \eqref{jT} into Eq.~\eqref{anm2}
and evaluating the integrals with
$r_x=r\sin\theta\cos \varphi$,
$r_y=r\sin\theta\sin \varphi$,
and
$r_z=r\cos\theta$, 
we find beam-shape coefficient up to the quadrupole order,
\begin{subequations}
\label{anm_quadrupole}
    \begin{align}
        a_{00} &= \frac{\sqrt{4 \pi}}{p_0} 
        p_{\text{in}},\\
        a_{10} &= 2 \ii\sqrt{3\pi} \frac{  \rho_0 c_0 }{p_0} 
        v_{\text{in},z},\\
        a_{1,\pm 1} &= \ii\sqrt{6 \pi}  \frac{\rho_0 c_0 }{p_0}
        (\mp v_{\text{in},x} + \ii v_{\text{in},y}),  \\
        a_{20} &= -\ii \sqrt{5\pi}
        \frac{\rho_0 c_0 }{k p_0}
        \left(
        \partial_x v_{\text{in},x} + \partial_y v_{\text{in},y}
        -2 \partial_z v_{\text{in},z}\right),\\
        a_{2,\pm 1} &= \ii \sqrt{30\pi}
        \frac{\rho_0 c_0 }{k p_0}
        \left( \mp \partial_z v_{\text{in},x} +
        \ii \partial_z v_{\text{in},y}
        \right),\\
        a_{2,\pm 2} &= \ii\sqrt{\frac{15 \pi}{2}}
        \frac{\rho_0 c_0 }{k p_0}
        \left(
        \partial_x v_{\text{in},x} - \partial_y v_{\text{in},y} 
        \mp 2 \ii \partial_x v_{\text{in},y} 
        \right).
    \end{align}
\end{subequations}
In this derivation, we assumed  the fluid flow is irrotational--see Eq.~\eqref{irrot}.
We remark  the acoustic fields of \eqref{anm_quadrupole} are evaluated at
$\vec{r}=\vec{0}$.

\subsection{Acoustic radiation force}
As the linear momentum flux carried by a wave is exchanged 
with the particle, the radiation force (a time-average quantity over the wave period $2\pi/\omega$) appears on the particle~\cite{Lopes2016},
\begin{equation}
\label{RFdef}
\vec{F}^\text{rad} = -
\int_{S_1} \re\left[\left(\frac{\beta_0 |p|^2}{4} - \frac{\rho_0 |v|^2}{4} 
 \right) {\bf I} + \frac{\rho_0}{2} \vec{v}\vec{v}^*\right]\cdot \vec{e}_r
 \, \dd S,
\end{equation}
where $S_1$ is a spherical surface that encloses the particle,
$\bf I$ is the unit tensor given in Eq.~\eqref{unitTensor}, and $\dd S$ is the area element.
The combination of two vectors as $\vec{v}\vec{v}$ forms a dyad, i.e., a second-rank tensor--see details in Appendix~\ref{app:dyads}.
The total pressure and fluid velocity are $p=p_\inc+p_\text{sc}$ and $\vec{v}=\vec{v}_\inc+\vec{v}_\text{sc}$, respectively.

In the long-wavelength limit $\epsilon\ll1$, the Cartesian components of the acoustic radiation force exerted on a spheroidal particle are expressed by~\cite{Silva2018}
\begin{widetext}
	\begin{subequations}
		\label{radforce}
		\begin{align}
		\nonumber
		F_x + \ii
		F_y &=\frac{\ii E_0}{2k^2}  \biggl[\sqrt{\frac{2}{3}}\left[a_{00}^*a_{1,-1} \left(s_{00}^*+s_{11} + 2 s_{00}^* s_{11}\right) +a_{00} a_{11}^* 
		\left(s_{00} + s_{11}^* + 2 s_{11}^* s_{00}\right)\right]\\
		&
		+\sqrt{\frac{2}{5}}\left(a_{10} a_{21}^* s_{10} + a_{2,-1}a_{10}^* s_{10}^*+ \sqrt{2}[
		a_{11} a_{22}^* s_{11}+ a_{2,-2} a_{1,-1}^*s_{11}^*]\right)
		+\sqrt{\frac{2}{15}} \left(a_{20}^* s_{11} a_{1,-1}+a_{20} a_{11}^* s_{11}^*\right)\biggr]
		,\\
		F_z &=\frac{E_0}{k^2} \im
		\biggl[
		\frac{2}{\sqrt{15}}a_{10}a^*_{20}s_{10}+
		\frac{1}{\sqrt{5}}
		(a_{1,-1}a^*_{2,-1} + a_{11} a_{21}^*) s_{11}
		+
		\frac{1}{\sqrt{3}}a_{00}a_{10}^*(s_{00} +s_{10}^*+
		2s_{00}s^*_{10})
		\biggr],
		\end{align}
	\end{subequations}
\end{widetext}
where the asterisk denotes complex conjugation, and $E_0=\beta_0 p_0^2/2$ is the characteristic energy density of the incident wave.

To obtain the radiation force as a function of the incoming acoustic fields, we
replace the beam-shape coefficients from \eqref{radforce} by
\eqref{anm_quadrupole}.
We also use the Helmholtz equation for the incoming pressure
$\partial_i\partial_i p_\text{in} = -k^2 p_\text{in}$.
Details on the derivation can be seen 
in Appendix~\ref{App:RF}. 
Accordingly, the radiation force is given by
	\begin{align}
	\nonumber
	\vec{F}^\text{rad}
	&= -\pi a^3\re\biggl[\beta_0\left({\bf Q}_{\text{grad}}^*-\ii \epsilon^3 {\bf Q}_{\text{sca}}^*\right)\cdot \nabla p_\text{in}
	\\ \nonumber
	&+ 
	\rho_0\left(\vec{D}_\text{grad}^*-\ii \epsilon^3\vec{D}_\text{sca}^*\right)\cdot \nabla \vec{v}_\text{in} 
	\biggr]_{\vec{r}=\vec{0}}, \\
	\label{Frad3}
	&= \vec{F}^\text{sca} + \vec{F}^\text{grad}.
	\end{align}
The scattering force results from the self-interaction of the scattering wave,
whereas the gradient force comes from the incoming and scattering wave interaction. 
Furthermore, the scattering force is much weaker than the gradient one by a factor of 
$\epsilon^3$.
The monopole tensors (${\bf Q}_\grad$ and ${\bf Q}_\text{sca}$) and dipole vectors ($\vec{D}_\text{grad}$ and $\vec{D}_\text{sca}$) are expressed by
\begin{subequations}
	\label{QD}
	\begin{align}
	{\bf Q}_\grad&= \frac{2}{3}f_{00} p_\text{in}(0) \,\vec{e}_i\vec{e}_i,\\
	\nonumber
	{\bf Q}_\text{sca}&= -\frac{ f_{00}}{9} p_\text{in}(0)
	\biggl[
	(2f_{00}+f_{11})\left(\vec{e}_x\vec{e}_x +\vec{e}_y\vec{e}_y\right)\\
	&+ 2(f_{00}+f_{10}) \,\vec{e}_z\vec{e}_z
	\biggr],\\
	\vec{D}_\text{grad}&=  -\frac{f_{11}}{2} \left[v_{\text{in},x}(0)\vec{e}_x
	+ v_{\text{in},y}(0)\vec{e}_y\right] - f_{10} v_{\text{in},z}(0)\vec{e}_z,\\
	\nonumber
	\vec{D}_\text{sca}&=  -\frac{1}{6}\\
	&
	\left[\frac{f_{11}^2}{4} \left[v_{\text{in},x}(0)\vec{e}_x
	+ v_{\text{in},y}(0)\vec{e}_y\right] + f_{10}^2 v_{\text{in},z}(0)\vec{e}_z\right].
	\end{align}
\end{subequations}
Note that $\vec{e}_i\vec{e}_j$ is a dyad, which forms the Cartesian basis of a second-rank tensor.
It is straightforward to show from \eqref{Frad3} that the gradient force is expressed by
\begin{subequations}
	\label{Fgrad}
	\begin{align}
	\vec{F}^\text{grad} &= - \nabla U(0),\\
	\nonumber
	U&=\pi a^3 \biggl[\frac{\beta_0 f_{00}}{3} |p_\text{in}|^2 - \frac{\rho_0}{2} \biggl(\frac{f_{11}}{2}(|v_{\text{in},x}|^2 + |v_{\text{in},y}|^2)\\
	& + f_{10} |v_{\text{in},z}|^2 \biggl)\biggr].
	\end{align}
\end{subequations}

Considering an  standing wave, the corresponding amplitude $p_\text{in}$ is a real-valued function.
Consequently, $\nabla p_\text{in}$ and ${\bf Q}^*_\text{sca}$ are also real-valued quantities. 
We thus conclude that 
$\re[\ii{\bf Q}^*_\text{sca}\cdot \nabla p_\text{in}] = \vec{0}$ and
$\re[\ii \vec{D}^*_\text{sca}\cdot \nabla \vec{v}_\text{in}] = \vec{0}$, so only the gradient force remains.

On the contrary, for traveling waves, which possess complex amplitude, say, a plane wave $\ee^{\ii \vec{k}\cdot \vec{r}}$,
the gradient operator becomes $\nabla \rightarrow \ii \vec{k}$, then $\vec{F}^\text{grad}=\vec{0}$.

\subsection{Acoustic radiation torque}
The acoustic radiation torque exerted on the particle is given by~\cite{Maidanik1958}
\begin{equation}
\vec{\tau}^\text{rad} =-\frac{1}{2}\re \int_{S_1} 
\vec{r}\times \rho_0 \vec{v}\vec{v}^*
\cdot \vec{e}_r\, \dd S.
\label{RTdef}
\end{equation}
The Cartesian components of the  radiation torque on a prolate spheroidal particle are given
by~\cite{Lopes2020,Leao-Neto2020}
\begin{subequations}
	\label{torque_rayleigh}
	\begin{align}
	\nonumber
	{\tau}_{x} &= - \frac{E_0 }{k^3\sqrt{2}}\,\text{Re} 
	\biggl[
	(a_{1,-1}+a_{11})(1+s_{11})a^*_{10}s^*_{10}\\
	&+
	(a_{1,-1}^* + a_{11}^*)(1+s_{10})a_{10}
	s_{11}^*
	\biggr],\\
	\nonumber
	{\tau}_{y} &=   \frac{E_0 }{k^3\sqrt{2}}\,\text{Re} \biggl[\ii\,
	(a_{1,-1}-a_{11})(1+s_{11})a^*_{10}s^*_{10}\\
	&-\ii\,
	(a_{1,-1}^* -a_{11}^*)
	(1+s_{10})a_{10}
	s_{11}^*
	\biggr],\\
	{\tau}_{z} &=  \frac{E_0 }{k^3 }\,\text{Re}
	\left[(|a_{1,-1}|^2 - |a_{11}|^2)
	(1+s_{11})s_{11}^*
	\right].
	\end{align}
\end{subequations}

We shall obtain the radiation torque in terms of the 
 momentum flux tensor and the acoustic spin density~\cite{Shi2019,Bliokh2019} of
the incident wave.
Here, they are expressed, respectively, by
\begin{subequations}
	\begin{align}
	\label{LMFlux}
	{\bf P}&=\frac{\rho_0}{2} \re[\vec{v}_\text{in}\vec{v}_\text{in}^*],\\
	\label{PS}
	{\bm S} &= \frac{\rho_0}{2 \omega}
	\im[\vec{v}_\text{in}^*\times\vec{v}_\text{in}].
	\end{align}
\end{subequations}
The acoustic spin is an intrisic property of a wave. It gives a measure of the fluid velocity rotation, which is independent of the reference frame.

To obtain the acoustic radiation torque, we substitute the scattering coefficients of \eqref{scatt_spheroid}
into \eqref{torque_rayleigh}.
The whole derivation is developed in Appendix~\ref{app:torque}.
Accordingly, the radiation torque is
\begin{equation}
\vec{\tau}^\text{rad}
= -\pi a^3 \biggl[   \chi 
\left[\vec{e}_z\times {\bf P}({0})\cdot \vec{e}_z\right] -
\frac{\epsilon^3}{24} \chi^2   \omega    \vec{S}_\perp({0})\biggr],
\label{torque_rayleigh4}	
\end{equation}
where $\chi=f_{11}- 2 f_{10} $ is the  dipole-difference  factor
and  $\vec{S}_\perp = S_x \vec{e}_x + S_y \vec{e}_y$ is the transverse spin density.
It is worth noticing that
$(  {\bf P}\cdot \vec{e}_z)_i= (\rho_0/2)\re[v_{\text{in},i}v_{\text{in},z}^*]$.

The first term of Eq.~\eqref{torque_rayleigh4} is related to the moment arm caused
the linear momentum flux ${\bf P}$
applied along the axial direction ($z$ axis).
This contribution  does not depend on the frequency.
The second term is the spin-induced torque,
which is much weaker than the momentum arm term.
It also varies with the frequency to the third power $\omega^3$.

\begin{figure}
	\includegraphics[scale=.32]{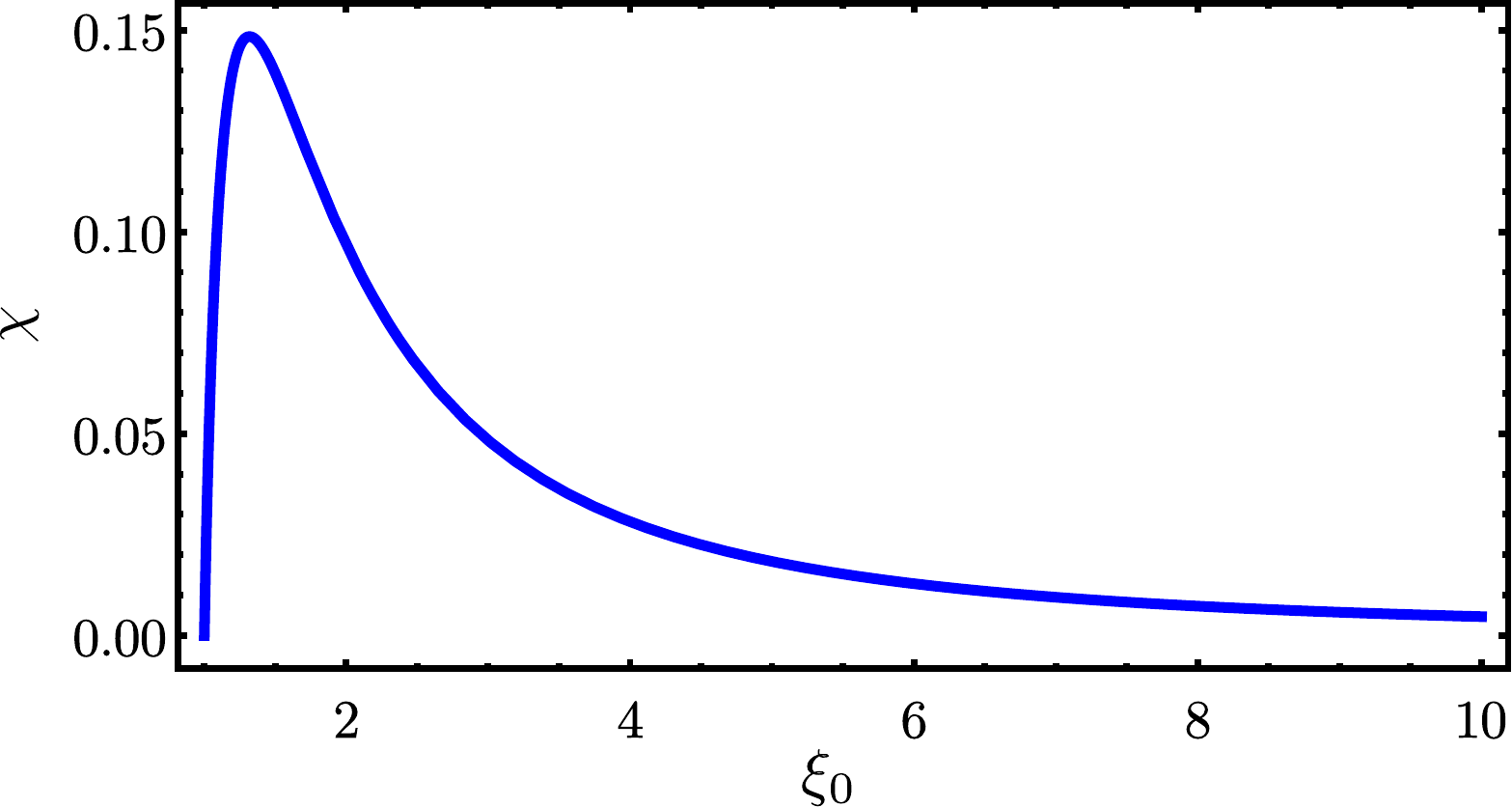}
	\caption{
		The dipole-difference factor $\chi$ is plotted as a function of
		the geometric parameter of the particle.
		The  maximum  value occurs at $\xi_0=1.3181$.
		\label{fig:tau1}}
\end{figure}

According to \eqref{f_sphere},
as the particle approaches an isotropic geometry ($\xi_0\rightarrow \infty$),
the  dipole-difference factor becomes
$\chi=0$. 
Thus, the radiation torque vanishes, as predicted before~\cite{Silva2012,Zhang2013}.
In Fig.~\ref{fig:tau1}, we show that the dipole-difference factor $\chi$ for a rigid spheroid. 
It is always positive and has a maximum at $\xi_0=1.3181$.
Besides, $\chi$ goes to zero as the particle approaches a thin line, e.g., $\xi_0\rightarrow 1$.
Thus,
the radiation torque  vanishes in this limit.

\section{Results and discussion}
\subsection{Two crossed plane waves}
\label{sec:2CPW}
\begin{figure}
	\includegraphics[scale=.5]{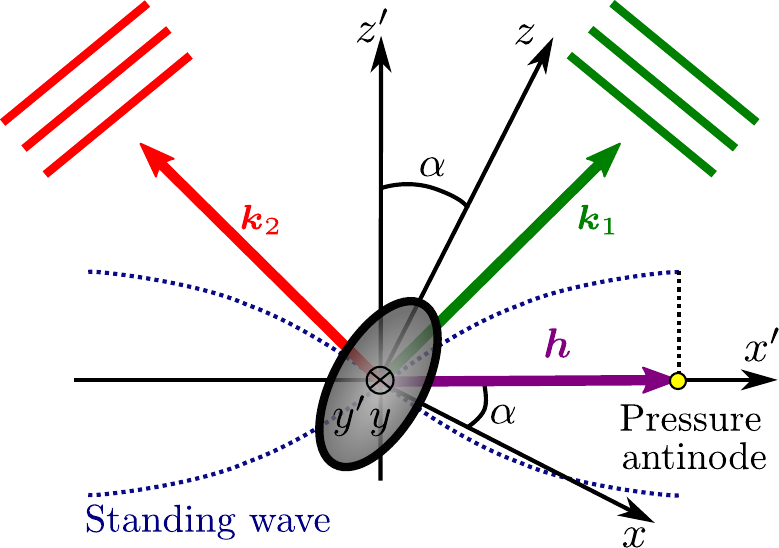}
	\caption{
		Two plane waves crossing at right angle
		and
		interacting with a spheroidal particle in the $x'z'$ plane.
		The particle frame corresponds to the $x$ and $z$ axes. 
		The angle $\alpha$ gives the particle orientation with respect to the $z'$-axis.
		The $y'$- and $y$-axis are going into the screen.
		The wavevectors $\vec{k}_1$ and $\vec{k}_2$ 
		are depicted by red and green arrows.
		The vector $\vec{h}$ (purple arrow) points to a transverse pressure antinode.
		\label{fig:2PlaneWaves}}
\end{figure}

Now we applied the developed theory to study the interaction of a subwavelength spheroid with  two crossed plane waves
at right angle.
This acoustic beam  was chosen because it has  acoustic spin~\cite{Shi2019}.
Moreover, as we shall see later, this beam forms simultaneously a traveling and standing wave.
In Fig.~\ref{fig:2PlaneWaves}, we sketch wave-particle interaction.
The wave vectors in the laboratory frame are
\begin{subequations}
	\label{WVlab}
	\begin{align}
	\vec{k}'_1&=\frac{k\sqrt{2}}{2} \left(  \vec{e}_{x'} +  \vec{e}_{z'}\right),\\
	\vec{k}'_2&=\frac{k\sqrt{2}}{2} \left(  -\vec{e}_{x'} +  \vec{e}_{z'}\right).
	\end{align}
\end{subequations}
Hence,
the corresponding pressure field is 
\begin{align}
\nonumber
p_\text{in} &=
\frac{p_{0}}{2} 
\left[ \ee^{\ii\vec{k}'_1\cdot (\vec{r}' + \vec{h}') } + \ee^{\ii\vec{k}'_2\cdot (\vec{r}' + \vec{h}') }
\right],\\
&=p_0 \cos \left[\frac{k}{\sqrt{2}} (x'+h) \right]\ee^{\ii k z'}.
\label{phi_cross0}
\end{align}
We see that the wave interference sets in
a standing wave along  the transverse direction ($x'$ axis).
The offset vector $\vec{h}'=h \vec{e}_{x'}$  gives the position of the nearest pressure antinode
regarding the particle center in the transverse direction.

To calculate the wave vectors in the particle system,
we introduce the (clockwise) rotation matrix around the $y'$ axis, 
\begin{equation}
\label{Ry}
{\bf R}_{y'}(\alpha)=
\left(
\begin{matrix}
\cos \alpha & 0 & -\sin \alpha\\
0 & 1 & 0\\
\sin \alpha & 0 &\cos \alpha
\end{matrix}
\right).
\end{equation}
From \eqref{WVlab}, we obtain the 
wave vectors in the particle system  as
\begin{subequations}
	\begin{align}
	\nonumber
	\vec{k}_1&= {\bf R}_{y'}(\alpha)\,\vec{k}'_1\\
			&=		
	\frac{k}{\sqrt{2}}\left[\left(\cos \alpha - \sin \alpha\right)\vec{e}_x + \left(\cos \alpha + \sin \alpha\right)\vec{e}_z\right],\\
	\nonumber
	\vec{k}_2 &= {\bf R}_{y'}(\alpha)\,\vec{k}'_2\\
	&=-\frac{k}{\sqrt{2}}\left[\left(\cos \alpha + \sin \alpha\right)\vec{e}_x - \left(\cos \alpha - \sin \alpha\right)\vec{e}_z\right].
	\end{align}
\end{subequations}
Also, the offset parameter becomes
$\vec{h}=h (\cos \alpha\,\vec{e}_x + \sin\alpha\,\vec{e}_z)$.
We thus obtain the incident pressure in the particle frame,
\begin{align}
\nonumber
p_\text{in} &=
\frac{p_{0}}{2} 
\left[ \ee^{\ii \vec{k}_1\cdot (\vec{r} + \vec{h}) } + \ee^{\ii\vec{k}_2\cdot (\vec{r} + \vec{h}) }
\right],\\
\nonumber
&=
\frac{p_{0}}{2} 
\biggl[ \ee^{\ii k[(x + z) \cos \alpha - (x-z)\sin\alpha +h]/\sqrt{2}} \\
&+ \ee^{-\ii k[(x - z ) \cos \alpha + (x+z)\sin\alpha + h]/\sqrt{2}}
\biggr].
\label{phiO}
\end{align}
The related fluid velocity is readily calculated by
inserting this equation into Eq.~\eqref{euler},
\begin{equation}
\vec{v}_{\text{in}} =\frac{  p_{0} }{2\rho_0 c_0 k }
\biggl[\vec{k}_1
\ee^{\ii \vec{k}_1\cdot (\vec{r} + \vec{h}) }+ \vec{k}_2 \ee^{\ii\vec{k}_2\cdot (\vec{r} + \vec{h}) }\biggl].
\label{v_cross}
\end{equation}

The linear momentum flux 
of the incoming beam is derived by replacing Eq.~\eqref{v_cross} into \eqref{LMFlux}, 
\begin{align}
\nonumber
{\bf P}(0) &= \frac{E_0}{4}\re\biggl[\vec{k}_1 \vec{k}_1 + \vec{k}_2\vec{k}_2 +\vec{k}_1\vec{k}_2 \ee^{\ii (\vec{k}_1-\vec{k}_2)\cdot \vec{h}}\\
&+ \vec{k}_2\vec{k}_1\ee^{-\ii (\vec{k}_1-\vec{k}_2)\cdot \vec{h}}\biggr].
\label{P_cross}
\end{align}
But $\vec{k}_1-\vec{k}_2 = \sqrt{2} k \, (\cos \alpha \,\vec{e}_x + \sin \alpha\, \vec{e}_z)$, then
\begin{equation}
{\bf P}(0) = \frac{E_0}{4}\re\biggl[\vec{k}_1 \vec{k}_1 + \vec{k}_2\vec{k}_2 +\vec{k}_1\vec{k}_2 \ee^{\ii \sqrt{2} k h}+ \vec{k}_2\vec{k}_1\ee^{-\ii \sqrt{2} k h}\biggr].
\end{equation}
The acoustic spin density  is obtained by substituting Eq.~\eqref{v_cross} into \eqref{PS},
\begin{align}
\nonumber
\vec{S}(0) &= -\frac{E_0}{4 \omega k^2} \im \left[(\vec{k}_2 \times \vec{k}_1)
\left(\ee^{-\ii (\vec{k}_1-\vec{k}_2)\cdot \vec{h}} -
\ee^{\ii (\vec{k}_1-\vec{k}_2)\cdot \vec{h}}
\right)\right]\\
&= \frac{E_0}{2 \omega}  \sin \left(\sqrt{2} k h\right) \vec{e}_y=\vec{S}_\perp(0).
\label{S_cross}
\end{align}
Here we have utilized the relation  $\vec{k}_2 \times \vec{k}_1 = k^2\, \vec{e}_y$.

\subsection{Radiation torque}
\label{sec:radtorque}
%
%

The  radiation torque in the laboratory frame is derived by 
substituting Eqs.~\eqref{P_cross} and \eqref{S_cross}  into Eq.~\eqref{torque_rayleigh4}, and noting that
$\vec{e}_y=\vec{e}_{y'}$.
Thus, we obtain
\begin{align}
\nonumber
&\vec{\tau}^\text{rad}
=\\
& \frac{\pi a^3 E_0 }{4}\biggl[\chi 
\cos\left(\sqrt{2}kh \right) \sin 2 \alpha
+\frac{ \epsilon^3}{12} \chi^2\sin\left(\sqrt{2} k h \right)
\biggr]\vec{e}_{y'}.
\label{torque_crossed}
\end{align}
When the particle is trapped in a pressure node
($h=\pi/k\sqrt{2}$) or antinode ($h=0$), the radiation torque becomes
\begin{equation}
\vec{\tau}^\text{rad}_\text{node}= - \vec{\tau}^\text{rad}_\text{anti} =-
 \frac{\pi a^3  }{4}E_0 \chi 
 \sin 2 \alpha\,\vec{e}_{y'}.
 \label{radtorque}
\end{equation}
These torques are caused exclusively by the momentum arm effect.
At the end, the particle will be aligned to broadside orientation $(\alpha=0)$ in a pressure node, and parallel  to the standing wave axis $(\alpha=\pi/2)$ in a pressure antinode.

Another interesting field position is $h=h_\pm= \pm \sqrt{2}\pi/4k$.
In this case, only the spin-induced torque arises,
\begin{equation}
 \vec{\tau}^\text{rad}_\text{spin} (h_\pm) =\pm \frac{\epsilon^3}{48} \pi a^3 E_0 \chi^2 \vec{e}_{y'}.
\end{equation}
This torque  sets the particle to rotate around the minor axis.
The particle spins in clockwise direction at $h=h_+$ (spin up), and in counterclockwise manner at $h=h_-$ (spin down).

\begin{figure}
	\includegraphics[scale=.47]{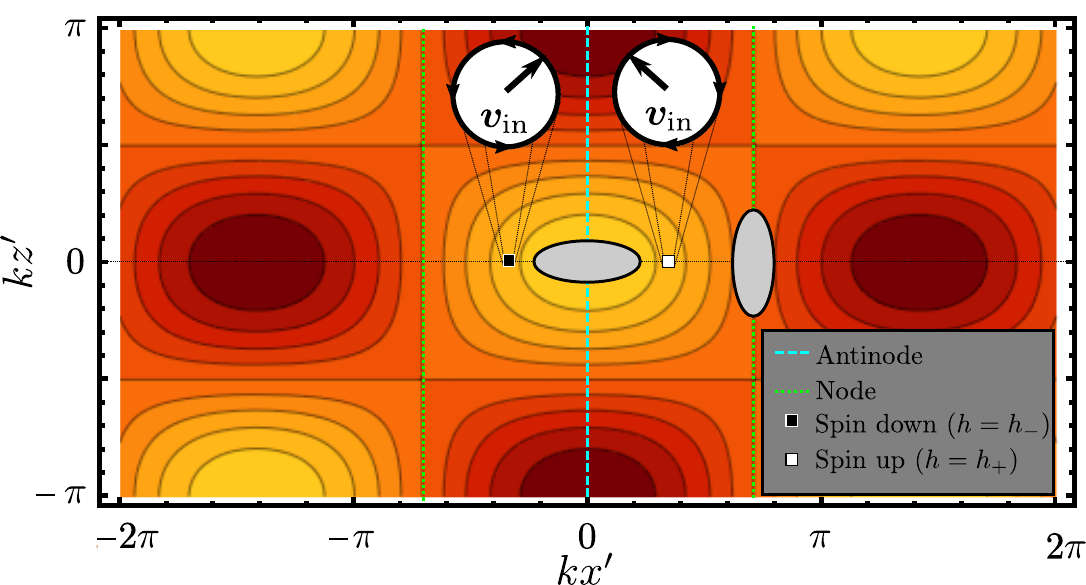}
	\caption{
		The amplitude (real part) of the incoming wave in Eq.~\eqref{phi_cross0}.
		The equilibrium orientation of the spheroidal particle (gray ellipses) are illustrated in a node ($\alpha=0$) and antinode ($\alpha=\pi/2$).
		The fluid velocity polarizabilities are shown when $h=h_+ =  \sqrt{2}\pi/4k$ (spin up) and  $h=h_-=-\sqrt{2}\pi/4k$ (spin down).
		Color legend: $p_0$ (bright yellow) and $-p_0$ (dark red).
		\label{fig:crossedwaves}
	}
\end{figure}

The incoming beam alongside the particle orientations in a node and antinode are illustrated in Fig.~\ref{fig:crossedwaves}.
The local fluid velocity polarizabilities are also shown for $h=h_\pm$. 
In this case, the velocities rotate circularly.

\subsection{Radiation force}
The transformation that furnishes the radiation force in the laboratory frame is 
\begin{equation}
\label{RFxlab}
{\vec{F}^\text{rad}}' = {\bf R}_{y'}(-\alpha) \vec{F}^\text{rad}.
\end{equation}
Here the matrix ${\bf R}_{y'}(-\alpha)$ defined in Eq.~\eqref{Ry} represents a counterclockwise rotation by an angle $\alpha$ around the $y'$-axis.

In the particle frame, the gradient force is derived
by substituting  Eqs.~\eqref{phiO} and \eqref{v_cross} into \eqref{Fgrad}.
After some straightforward calculations, we arrive at 
	\begin{align}
	\nonumber
	\vec{F}^\text{grad}
	&= \frac{{\epsilon}}{12\sqrt{2}}\pi a^2  E_0\left[ 8f_{00} + 3( f_{11} -2 f_{10}) \cos 2\alpha \right]\\
	& \sin\left(\sqrt{2} k h \right)
	\left(\cos \alpha\, \vec{e}_x + \sin \alpha\, \vec{e}_z\right).
		\label{RFparticle}
	\end{align}
Inserting Eq.~\eqref{RFparticle} into \eqref{RFxlab}, we obtain the transverse radiation force in the laboratory frame,
\begin{subequations}
	\label{trans_RF_lab}
	\begin{align}
		{\vec{F}^\text{grad}}'
		 &= \epsilon \pi a^2 E_0   \Phi_\text{ac}
  		\sin(\sqrt{2} k h )\, \vec{e}_{x'},\\ 
   		\Phi_\text{ac} &=\frac{1}{12\sqrt{2}} \left[ 8f_{00} + 3( f_{11} -2 f_{10}) \cos 2\alpha\right].
  		\label{AC}
  		\end{align}
\end{subequations}
The sign of the  acoustophoretic factor $\Phi_\text{ac}$ determines whether the particle will be  trapped in a pressure node ($\Phi_\text{ac}>0$) or antinode ($\Phi_\text{ac}<0$).
Moreover, the  radiation force varies linearly with frequency, $\epsilon\sim \omega$.
We will analyze to the axial radiation force later.

\begin{figure}
	\includegraphics[scale=.32]{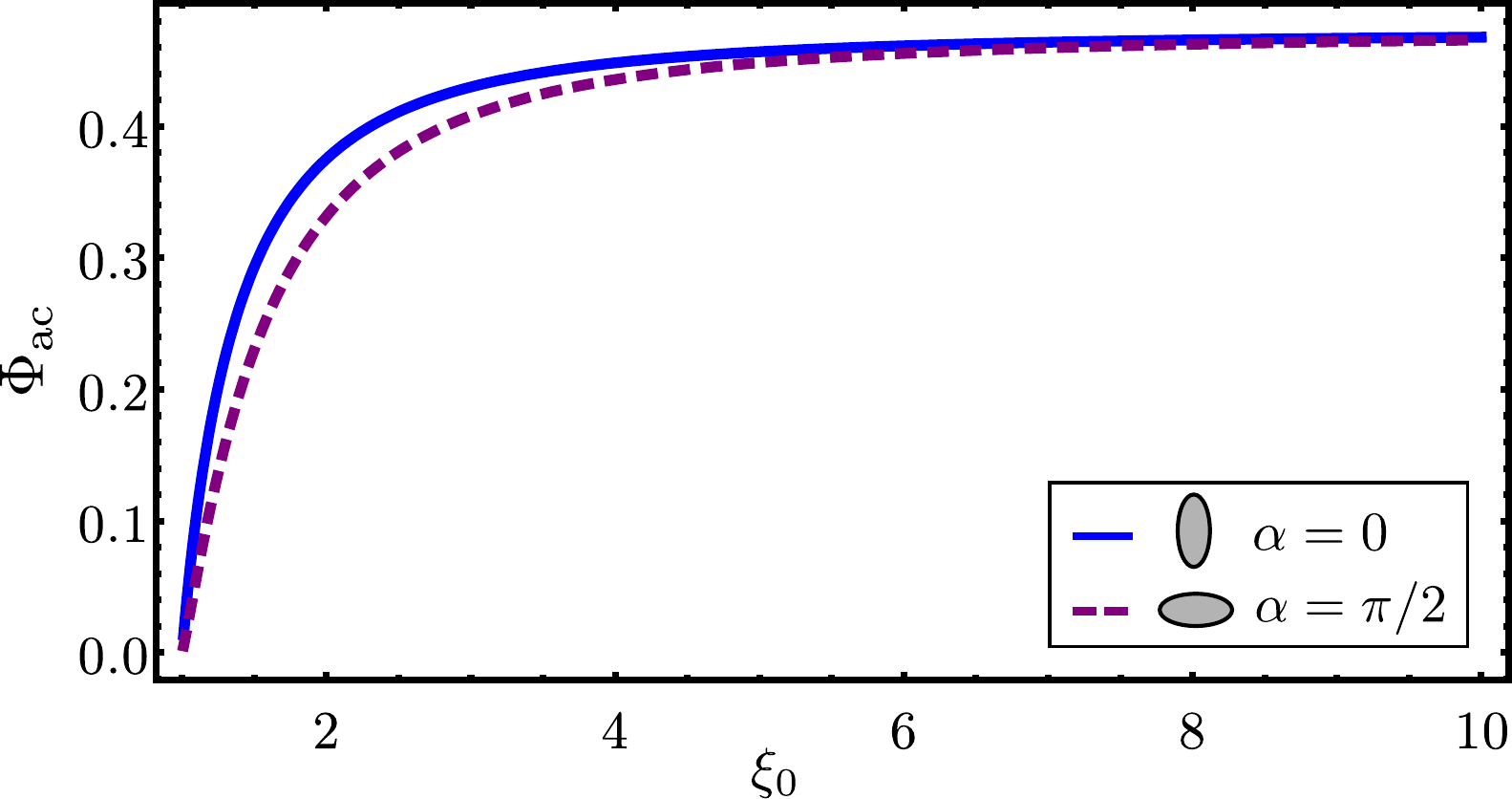}
	\caption{
		The acoustophoretic factor $\Phi_\text{ac}$ versus  the geometric parameter $\xi_0$ for different particle orientations. 
		\label{fig:Phi_ac}}
\end{figure}

We plot the acoustophoretic factor $\Phi_\text{ac}$ versus $\xi_0$ in Fig.~\ref{fig:Phi_ac}.
No significant difference is noted as the orientation
changes from $0$ to $\pi/2$.
The factor is positive; thus, the particle will be transversely trapped in a pressure node.
As the particle approaches a spherical shape $\xi_0\rightarrow \infty$, then $\Phi_\text{ac}\rightarrow \sqrt{2}/3$.
In the other extreme, as $\xi_0\rightarrow 1$, we have
$\Phi_\text{ac}\rightarrow 0$.
Hence, no transverse radiation force is produced on a thin line particle.

Now we take a closer look at the axial radiation force
as the particle is trapped in a  pressure node.
This force is in fact the scattering force defined in \eqref{Frad3}.
Again, we insert Eqs.~\eqref{phiO} and \eqref{v_cross} into \eqref{Frad3} and set $h=\pi/\sqrt{2} k$ (pressure node) to find 
\begin{subequations}
	\label{RFe6}
	\begin{align}
	F_x^\text{sca} &=
-\frac{\pi a^2 \epsilon^4 E_0}{24 \sqrt{2} }
\left[
f_{11}^2 + ( f_{11}^2 - 4 f_{10}^2 )\cos 2\alpha
\right]
\sin \alpha,\\
F_z^\text{sca} &=\frac{\pi a^2 \epsilon^4 E_0}{24 \sqrt{2} }
\left[4
f_{10}^2 + (f_{11}^2 - 4 f_{10}^2  )\cos 2\alpha
\right]
\cos \alpha.
\end{align}
\end{subequations}
Applying the rotation operator ${\bf R}_{y'}(-\alpha)$ on \eqref{RFe6} yields the axial radiation force in the laboratory frame,
\begin{subequations}
	\label{Fzp}
\begin{align}
\label{RFzlab6}
{\vec{F}^\text{sca}}' &= \pi a^2 
  E_0 Q_\text{rad} \, \vec{e}_{z'}, \quad \text{(pressure node),} \\
  \label{Qsca}
Q_\text{rad} &= \frac{\epsilon^4 }
{48\sqrt{2} } \left[
4 f_{10}^2 + f_{11}^2 + (f_{11}^2 - 4f_{10}^2)\cos 2\alpha
\right].
\end{align}
\end{subequations}
The quantity $Q_\text{rad}$ is referred to as
the dimensionless radiation force efficiency.
We remark  the scattering force is related to the linear momentum flux carried away by the scattered waves.
This connection can be understood from the radiation force relationship with the scattering cross-section as discussed in Refs.~\cite{Zhang2011,Leao-Neto2017}.
In turn, the scattering cross-section of a rigid spheroid scales as $\pi a^2(ka)^4$~\cite{Silbiger1963},
which is the same dependence seen in the scattering force of Eq.~\eqref{RFzlab6}.

\begin{figure}
	\includegraphics[scale=.32]{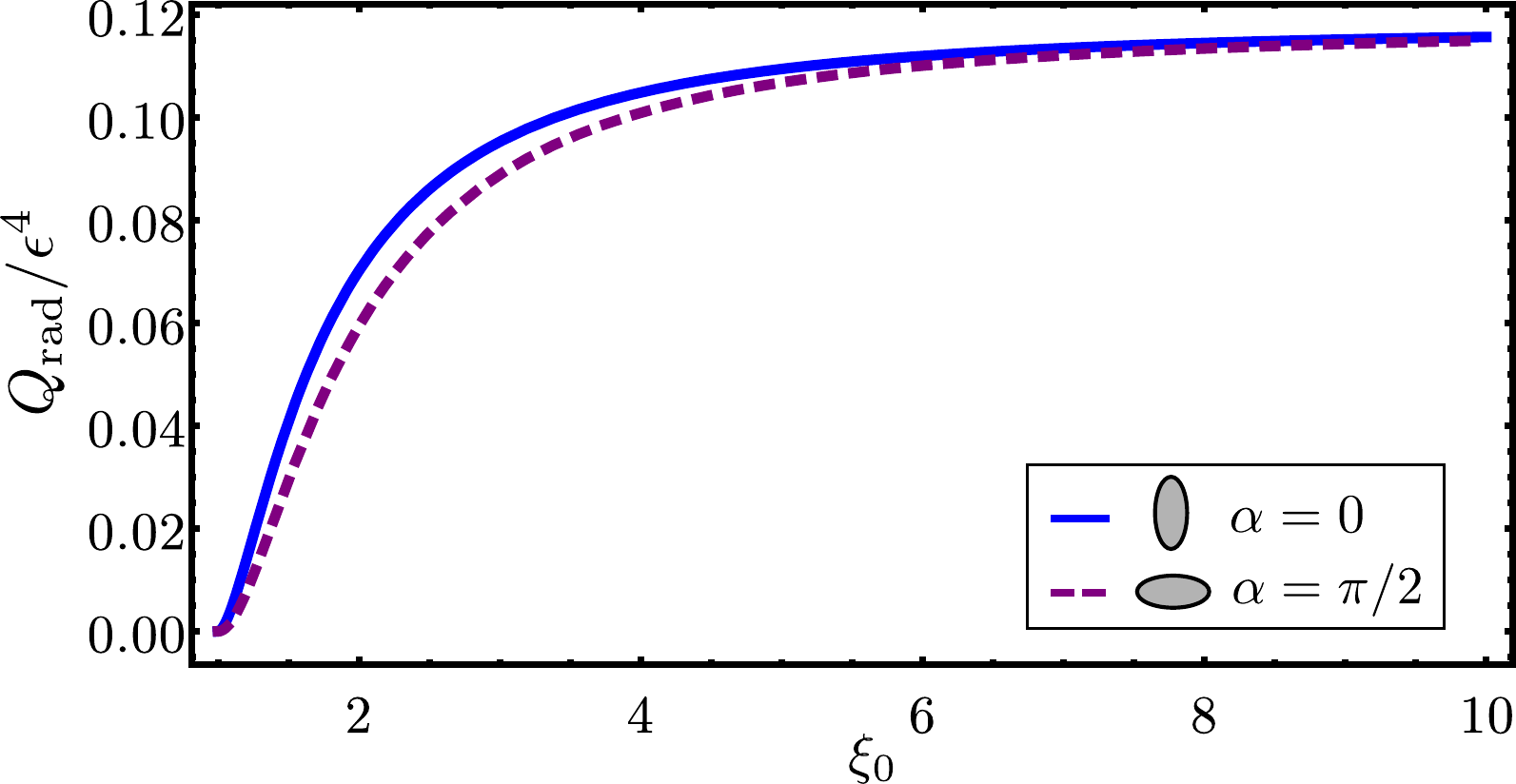}
	\caption{
		The radiation force efficiency $Q_\text{rad}$ normalized to $\epsilon^4$ versus the geometric parameter of the particle for different orientations.
		\label{fig:Qrad}}
\end{figure}

In Fig.~\ref{fig:Qrad}, we plot the radiation force efficiency divided by $(ka)^4$ versus with $\xi_0$.
The efficiency is positive and asymptotically approaches
$Q_\text{rad}/(ka)^4\sim 1/6\sqrt{2}\approx0.118$
as $\xi_0\rightarrow\infty$.
Also, it goes to zero as $\xi_0\rightarrow1$.
We see that $Q_\text{rad}$ is positive
with no significant difference as the orientation
changes from $0$ to $\pi/2$.
Thereby, the force pushes the particle along the direction of the traveling component of the incoming wave.

An important aspect of Eq.~\eqref{RFzlab6} is
the possibility 
to recover the axial radiation force imparted to a rigid spherical particle as obtained in Ref.~\cite{Xu2012}.
By replacing the scattering factors of \eqref{f_sphere} for a spherical particle of radius $r_0$ into \eqref{RFzlab6},
we encounter
$
F_{z'}^\text{sphere}=\pi r_0^2(k r_0)^4 E_0/6\sqrt{2}.
$
This result agrees with Ref.~\cite{Xu2012}.
To see this connection, we multiply Eq.~(15b) of~\cite{Xu2012} by $\pi r_0^2 E_0/4$, and
  set the parameters, in the paper's notation, to ${\rho}=1/2$, ${\kappa}=-1$ (rigid sphere) and $\gamma=\pi/4$.
\begin{table}
	\caption{\label{tab:parameters}
		Parameters used in the finite-element simulations in Comsol at room temperature and pressure.
	}
	{
		\begin{tabular}{lr}
			\hline
			\hline
			\textbf{Parameter} & \textbf{Value}\\
			\hline
			\textbf{Spheroidal particle}& \\
			Major semiaxis ($a$)   &    $\SI{47.14}{\micro\meter}$\\
			Minor semiaxis ($b$) & $\SI{30.71}{\micro\meter}$\\
			Geometric parameter ($\xi_0$)  &  $1.3181$\\
			Size parameter ($\epsilon$) & $0.2$\\
			\hline
			\textbf{Medium (water)} &\\
			Density ($\rho_0$) &  $ \SI{999.66}{\kilogram \per \meter\cubed}$\\
			Speed of sound ($c_0$) & $ \SI{1481}{\meter \per \second}$\\
			Compressibility ($\beta_0$) & $\SI{0.456}{\per\giga\pascal}$ \\
			Domain radius & $5a=\SI{235.7}{\micro\meter}$\\
			Radius of $S_1$& $a+b/4=\SI{54.82}{\micro\meter}$ \\
			Maximum element size inside $V_1$ & $b/18=\SI{1.706}{\micro\meter}$\\
			Maximum element size outside $V_1$  & $\lambda/48=\SI{30.71}{\micro\meter}$\\
			PML depth ($15$ layers)  & $2a=\SI{94.28}{\micro\meter}$ \\
			\hline
			\textbf{Acoustic wave} \\ 
			Pressure peak ($p_0$) & $\SI{100}{\kilo\pascal}$\\
			Acoustic energy density ($E_0$) & $\SI{2.28}{\joule\per\meter\cubed}$\\
			Linear frequency ($f$) & $\SI{1}{\mega\hertz}$\\
			Wavenumber ($k$) & $\SI{4242.5}{\per\meter}$\\
			Wavelength ($\lambda_0$) & $\SI{1.481}{\milli\meter}$\\
			\hline
			{\bf Computer system }& \\
			CPU  &       Xeon E5-2690 3.00GHz\\
			Operating system &        Linux\\
			Maximum memory usage & $\sim\SI{128}{\giga\byte}$\\
			Computational time per dataset & $\sim\SI{20}{\minute}$\\
			\hline
		\end{tabular}
	}
\end{table}
\begin{figure}
	\includegraphics[scale=.6]{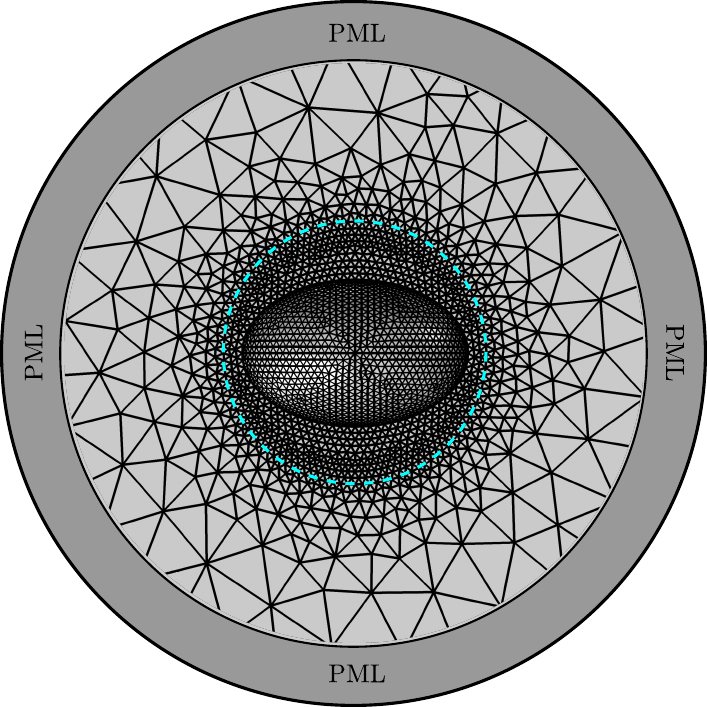}
	\caption{
		The sketch of the FE mesh used to compute the radiation force and torque on a spheroidal particle.
		The model comprises a spherical region as the simulation domain and the PML.
		The dashed-line (light blue) depicts the integration surface $S_1$. 
		The  particle is placed at the center of the simulation domain.
		A finer mesh with $1.746\times 10^6$ elements is defined inside the volume enclosed by $S_1$.
		Outside this region, a coarser mesh is used with $2.86\times 10^5$ elements.
		The PML has $25860$ elements.
		\label{fig:mesh}}
\end{figure}

\subsection{Finite-element simulations}
We performed FE simulations of 
the wave-particle interaction in water.
A set of full 3D simulations were devised in Comsol Multiphysics (Comsol, Inc., USA). 
The FE model used in our simulations is outlined as follows.
A spherical region of radius $R$ is defined as the simulation domain.
The particle is placed at the center of this region.
We set a spherical surface $S_1$ as the integration surface 
to compute the radiation force and torque as prescribed in Eqs.~\eqref{RFdef} and \eqref{RTdef}.
The volume between the particle surface $S_0$ and $S_1$ is denoted by $V_1$.
To mimic the Sommerfeld radiation condition given in Eq.~\eqref{Rcondition} for the scattered waves, we use
the perfect matched layer (PML). 
The rigid boundary condition for the particle as given in Eq~\eqref{velBoundary} is assumed.
The sketch of the FE model is displayed in Fig.~\ref{fig:mesh}. 
The incident beam is set as the background pressure in the particle frame.
The total pressure and fluid velocity fields are then computed and used to calculate the radiation force and torque.
The physical parameters were  inspired on those reported for acoustofluidic devices~\cite{Dron2012}.

To verify the correctness of the numerical solutions, we performed some convergence tests by varying some parameter such as the domain radius, mesh density, and PML depth.
The parameter variation is set 
up to the point the solution does not change with any further modification.
More details on convergence analysis to radiation force problems can be found in~\cite{Glynne-Jones2013}.
Typical accuracy achieved in our tests is less than $1\,\%$ for effects of order $\bigO(1)$ and $\bigO(\epsilon)$, 
and up to $10\,\%$ for $\bigO(\epsilon^3)$ and $\bigO(\epsilon^4)$.
The higher-order effects such as the spin-torque and scattering force seem to be more sensitive to reflections
on the PML.
The simulational parameters summarized in Table~\ref{tab:parameters} were chosen to warrant the aforementioned numerical accuracy.

To compare the obtained  numerical results with theory, we use the normalized root-mean-square error (NRMSE),
\begin{equation}
\text{NRMSE}\,(\%)= \frac{100}{x_\text{max} - x_\text{min}}\sqrt{\frac{1}{N}\sum_{n=1}^N \left(x_n^{\text{theory}}- x_n^{\text{num}}  \right)^2},
\end{equation}
in which $N$ is the number of sampling points,  $x_\text{max}=\max\{x_n^\text{theory}\} $
and $x_\text{min}=\min\{x_n^\text{theory}\}$.

Importantly, both the theoretical and numerical models consider the inviscid approximation.
Although, actual fluids may support shear stress within the particle boundary layer, which may cause viscous torques~\cite{Lee1989}.
When the boundary layer $\delta=(2\mu_0/\rho_0 \omega)^{1/2}$ ($\mu_0$ is the dynamic viscosity of the fluid), is
much smaller than the particle size, say $\delta/b\ll 1$, viscous torque can be discarded.
In our simulations, $\delta/b=10^{-2}$, and thus, we may neglect viscous torque effects here. 

\begin{figure}
	\includegraphics[scale=.495]{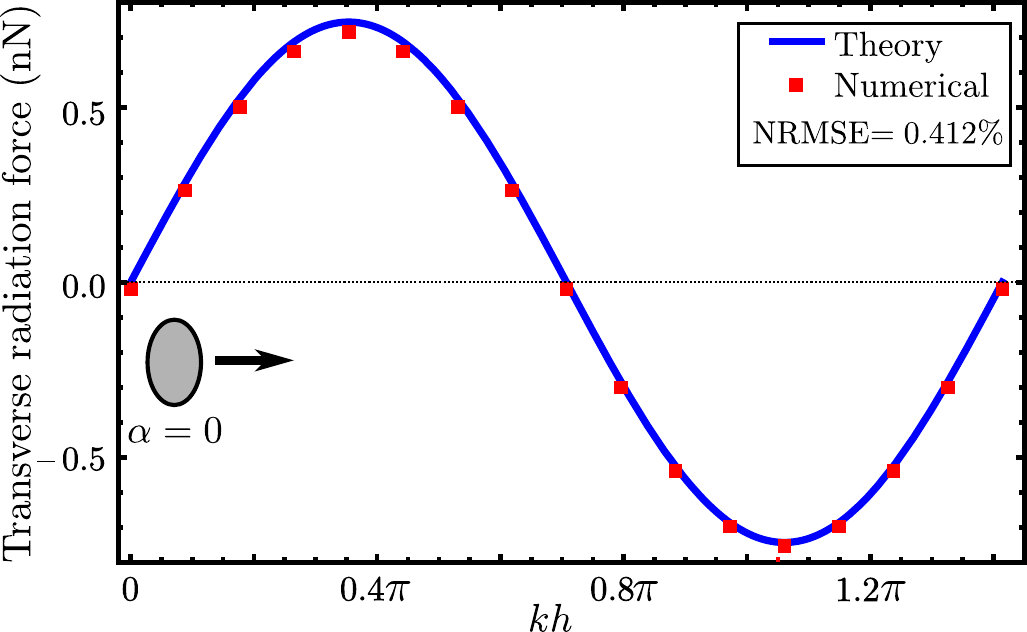}
	\caption{
		Numerical and theoretical calculations of the transverse radiation force (${{F}_{x'}^\text{grad}}$) on a spheroidal particle as a function of the scaled distance $kh$. 
		The particle (gray ellipse) has a geometric parameter of $\xi_0=1.3181$ with orientation $\alpha=0$.
		The simulational parameters are presented in Table~\ref{tab:parameters}.
		\label{fig:TransverseRF}}
\end{figure}

In Fig.~\ref{fig:TransverseRF}, we plot
the transverse radiation force $F^\grad_{x'}$ as a function of the scaled distance $kh$.
The particle (gray ellipse) has a geometric parameter of $\xi_0=1.3181$, which corresponds to the maximum amplitude of the dipole-difference factor seen in Fig.~\ref{fig:tau1}, and consequently the maximum of the radiation torque in Eq~\eqref{torque_crossed}. 
The radiation force is evaluated with the equations of
\eqref{trans_RF_lab}.
We see the particle will be trapped in the pressure node
$kh=\pi/\sqrt{2}=0.71\pi$.
Both visual inspection and $\text{NRMSE}=0.412\%$ indicate  the theoretical result is in remarkable agreement with the FE data.
\begin{figure}
	\includegraphics[scale=.48]{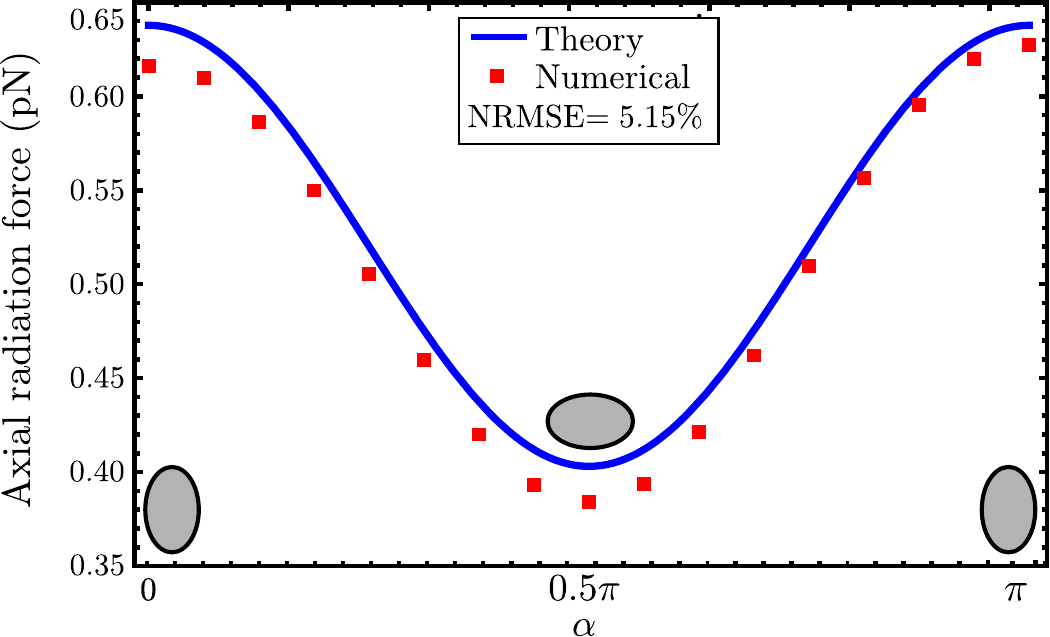}
	\caption{The axial radiation force (${{F}_{z'}^\text{sca}}$) at a pressure node versus  particle orientation.
		The particle (gray ellipses) has a geometric parameter of $\xi_0=1.3181$.
		The simulational parameters are surveyed in Table~\ref{tab:parameters}.
		\label{fig:RFaxial}}
\end{figure}

The axial radiation force ($F_{z'}^\text{sca}$) in a pressure node versus particle orientation is depicted in Fig.~\ref{fig:RFaxial}.
This force was calculated with the equations of~\eqref{Fzp}.
The $\text{NRMSE}=5.15\%$ is one order of magnitude above the error of the transverse force in Fig.~\ref{fig:TransverseRF}.
This is so because the scattering force is more sensitive to numerical errors, since it is much weaker than the transverse counterpart. 
The maximum  and minimum force is experienced by the particle as its orientation is $\alpha=0,\pi/2$, respectively.

\begin{figure}[t]
	\includegraphics[scale=.49]{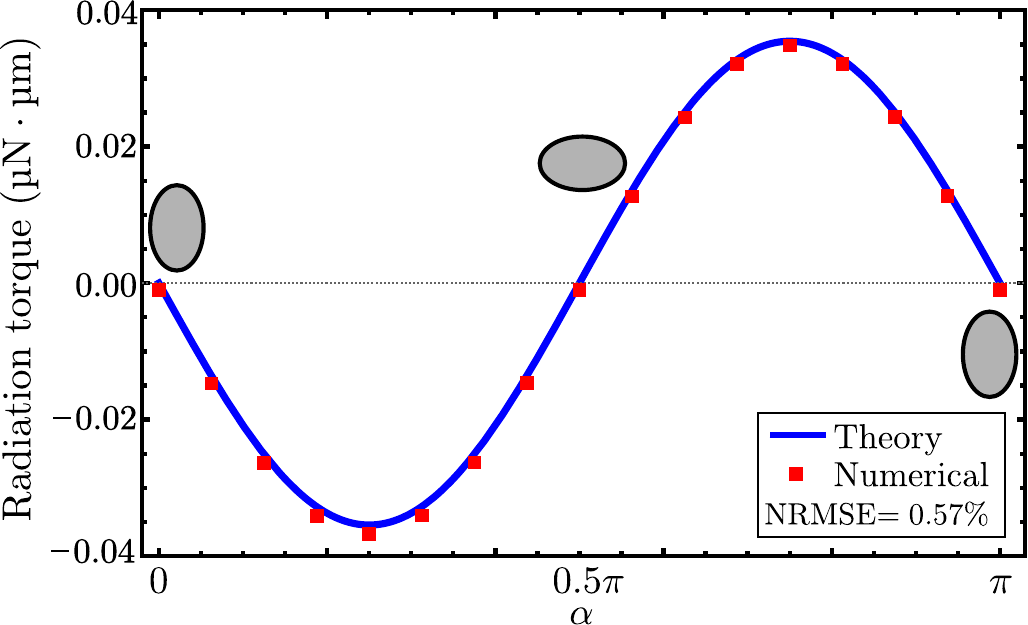}
	\caption{
		The theoretical and numerical results of
		the  radiation torque versus  orientation angle $\alpha$.
		The particle (gray ellipses) has a geometric parameter of $\xi_0=1.3181$ and is trapped in a pressure node.
		The simulational parameters are summarized in Table~\ref{tab:parameters}.
		\label{fig:RT}}
\end{figure}

In Fig.~\ref{fig:RT}, we show how the radiation torque varies with the particle orientation in a pressure node.
The torque is evaluated using Eq.~\eqref{radtorque}, and is caused by 
the momentum arm only.
As previously discussed, the preferential orientation of the particle is $\alpha=0$, i.e., the particle will be aligned perpendicularly to the standing wave axis.
Again, we find an excellent agreement between the theory and numerical results, provided that $\text{NRMSE}=0.57\%$.
\begin{figure}
	\includegraphics[scale=.48]{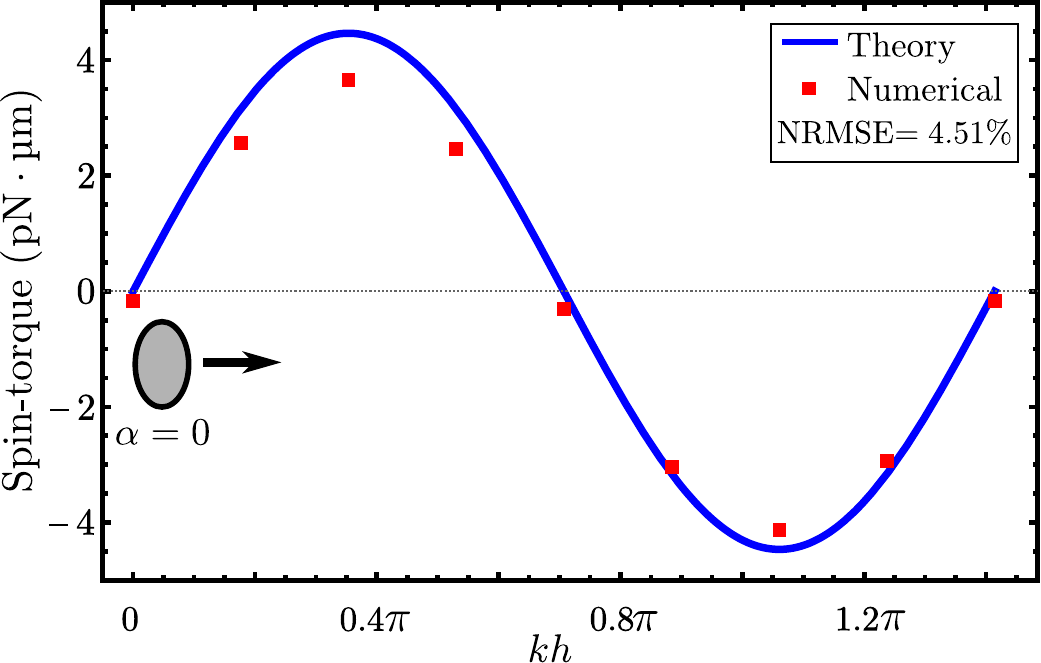}
	\caption{The
		theoretical and numerical  spin-torque on the spheroidal particle with a fixed orientation  $\alpha=0$. 
		The particle (gray ellipse) has a geometric parameter of $\xi_0=1.3181$ and
		its position varies from an antinode to a node.
		The simulational parameters are shown in Table~\ref{tab:parameters}.
		\label{fig:RTspin}}
\end{figure}

The amplitude of the spin-torque is shown in Fig.~\ref{fig:RTspin} versus the particle position $kh$.
The particle orientation is fixed to $\alpha=0$.
Equation~\eqref{torque_crossed} is used to calculate the radiation torque.
The theoretical and numerical data are in good match with $\text{NRMSE}=4.51\%$.
This error is ten times larger than that of the momentum arm torque in Fig.~\ref{fig:RT}.
Possibly, because the spin-torque is a weaker phenomenon and then is more sensitive to undesired wave reflections from the PML.

\section{Concluding remarks}
The acoustic radiation force and torque are the by-products of the nonlinear interaction of an incoming wave with a spheroidal particle.
We develop a theory to describe these phenomena considering subwavelength particles and an incoming wave of arbitrary character.
The main contribution of our work are the laconic expressions of the radiation force in Eqs.~\eqref{Frad3} and \eqref{Fgrad}, and radiation torque in Eq.~\eqref{torque_rayleigh4}.
In these expressions, the acoustic fields can be described either analytically, numerically, or experimentally.
Moreover, the particle anisotropy allows the rise of the radiation torque
from  the momentum arm and acoustic spin of the incoming wave.
Whereas, the radiation force is composed of the gradient and scattering forces similarly to the spherical particle case~\cite{Silva2014}.

Our theory is successfully employed to analyze the interaction between a rigid spheroidal particle and  two plane waves crossing at right angle.
This composed beam is one of the simplest waves to possess acoustic spin~\cite{Shi2019}.
When the particle is trapped in a pressure node, the induced radiation torque is solely due to the momentum arm caused by linear momentum flux.
However, as the particle is placed at $\pi\sqrt{2}/4k$ along the transverse direction (i.e., the $x$ axis), 
a clockwise spin-torque is activated.
The spin flips as the particle changes place to $-\pi\sqrt{2}/4k$.
Importantly, the spin-torque is  $(ka)^3$ weaker than the momentum arm contribution.
The most prominent effect of the acoustic radiation force is to trap the particle in a transverse pressure node.
The axial radiation force in a pressure node follows the traveling wave direction.
The theoretical results are verified against FE simulations  based on a full three-dimensional  model.
An excellent agreement is found between theory and numerical experiments.

The next level to be considered in our model is to bring in the
elastic properties of particles.
These features will convey the theory closer to acoustofluidic experiments with cells and other microorganisms.
Moreover, energy absorption by the particle can be as well considered through complex wave numbers.
The absorption is known to enhance the acoustic radiation force~\cite{Leao-Neto2016}.
With these extensions, it will be possible to predict the conditions to trap particles in either a pressure node or antinode.
Besides, devising a coated layer on spheroidal particles may render them unresponsive to the radiation force~\cite{Leao-Neto2016a}.
Also, getting to know how the elastic properties affect the radiation torque has great importance
to control the rotational degree of freedom in microparticle manipulation.

Our predictions can be tested with experiments of either acoustofluidics or acoustic levitation technology.
The radiation force and torque derived expressions are valid for any structured  wave that is commonly employed in these techniques.
Experimental confirmation of our results may unlock new methods for acoustic manipulation of particles.

The theoretical model of the nonlinear interaction between acoustic waves and anisotropic particles is a timely 
research contribution. 
It has a great potential within both acoustic manipulation of particles and the investigation of fundamental aspects of acoustic waves.

\begin{acknowledgments}
We thank the National Council for Scientific and Technological Development--CNPq,
Brazil (Grant No. 401751/2016-3 and No. 307221/2016-4) for financial support.
\end{acknowledgments}

\appendix
\section{Radiation force expressions}
\label{App:RF}

Inserting the beam-shape coefficients from \eqref{anm_quadrupole} into
\eqref{radforce}, using 
$\partial_i\partial_i p_\text{in} = -k^2 p_\text{in}$,
and rearranging the terms, we arrive at
\begin{subequations}
	\label{Frad}
	\begin{align}
	\left(
	\begin{matrix}
	F_x\\
	F_y\\
	F_z
	\end{matrix}
	\right)
	& = \re\left[
	\left(
	\begin{matrix}
	\mathcal{D}_{xy} & 0 & 0\\
	0 & \mathcal{D}_{xy} & 0\\
	0 & 0 & \mathcal{D}_{zz}
	\end{matrix}
	\right)
	\left(
	\begin{matrix}
	v_{\text{in},x}^*\\
	v_{\text{in},y}^*\\
	v_{\text{in},z}^*
	\end{matrix}
	\right)
	\right]_{\vec{r}=\vec{0}},\\
	\nonumber
	\mathcal{D}_{xy} &=-
	\frac{2\pi\ii }{k^2}
	\biggl[\frac{3\rho_0}{k}  \left(s_{11} 
	\vec{v}_\text{in}\cdot\nabla_\perp
	+s_{10}v_{\text{in},z}\partial_z\right)\\
	&-\ii
	s_{00}\left(1+2s_{11}^*\right)
	\frac{p_\text{in}}{ c_0 }
	\biggr],\\
	\nonumber
	\mathcal{D}_{zz} & =-
	\frac{2\pi \ii }{k^2}\biggl[
	\frac{3\rho_0}{k}  \left(s_{11} \vec{v}_\text{in}\cdot
	\nabla_\perp
	+s_{10}v_{\text{in},z}\partial_z\right)\\
	&
	-\ii
	s_{00}\left(1+2s_{10}^*\right)\frac{p_\text{in}}{ c_0 }
	\biggr].
	\end{align}
\end{subequations}
Replacing the scattering coefficients of \eqref{Frad} with \eqref{scatt_spheroid}, we obtain
\begin{subequations}
\begin{align}
	\nonumber
\vec{F}^\text{rad}
&= -\pi a^3\re\biggl[\beta_0\left({\bf Q}_{\text{grad}}^*-\ii \epsilon^3 {\bf Q}_{\text{sca}}^*\right)\cdot \nabla p_\text{in}
\\ \nonumber
&+ 
\rho_0\left(\vec{D}_\text{grad}^*-\ii \epsilon^3\vec{D}_\text{sca}^*\right)\cdot \nabla \vec{v}_\text{in} 
\biggr]_{\vec{r}=\vec{0}}, 
\label{Frad2}\\
	{\bf Q}_\grad&= \frac{2}{3}f_{00} p_\text{in}(0) \,\vec{e}_i\vec{e}_i,\\
\nonumber
{\bf Q}_\text{sca}&= -\frac{ f_{00}}{9} p_\text{in}(0)
\biggl[
(2f_{00}+f_{11})\left(\vec{e}_x\vec{e}_x +\vec{e}_y\vec{e}_y\right)\\
&+ 2(f_{00}+f_{10}) \,\vec{e}_z\vec{e}_z
\biggr],\\
\vec{D}_\text{grad}&=  -\frac{f_{11}}{2} \left[v_{\text{in},x}(0)\vec{e}_x
+ v_{\text{in},y}(0)\vec{e}_y\right] - f_{10} v_{\text{in},z}(0)\vec{e}_z,\\
\nonumber
\vec{D}_\text{sca}&=  -\frac{1}{6}\\
&
\left[\frac{f_{11}^2}{4} \left[v_{\text{in},x}(0)\vec{e}_x
+ v_{\text{in},y}(0)\vec{e}_y\right] + f_{10}^2 v_{\text{in},z}(0)\vec{e}_z\right].
\end{align}
\end{subequations}

\section{Radiation torque expressions}
\label{app:torque}
Substituting the beam-shape coefficients of \eqref{anm_quadrupole} into~\eqref{torque_rayleigh}, we obtain
\begin{subequations}
	\label{torque_rayleigh2}
	\begin{align}
	\left(
	\begin{matrix}
	{\tau}_{x}\\
	{\tau}_{y}
	\end{matrix}
	\right)
	& =  \frac{6 \pi \rho_0 }{k^3}\,\text{Im} 
	\left[
	(s^*_{10}+s_{11}+2s_{11}s^*_{10})
	\left(
	\begin{matrix}
	v_{\text{in},y} v_{\text{in},z}^*\\
	-v_{\text{in},x} v_{\text{in},z}^*
	\end{matrix}
	\right)
	\right],\\
	\label{tz}
	\tau_z &=  \frac{6 \pi \rho_0 }{k^3}
	\left(s_{11}+s_{11}^*+2|s_{11}|^2\right) \text{Im}\left[v_{\text{in},x} v_{\text{in},y}^*\right].
	\end{align}
\end{subequations}
After using the scattering coefficients of~\eqref{scatt_spheroid}
into Eq.~\eqref{tz}, we find  the axial radiation torque $\tau_z =  \bigO(\epsilon^{12})$.
This contribution is then neglected.

In contrast, the transverse components of the torque are given by
\begin{align}
\nonumber
\left(
\begin{matrix}
\tau_x\\
\tau_y
\end{matrix}
\right) &=
\frac{\pi {\epsilon}^3}{2k^3}   (f_{11}- 2 f_{10} )
\rho_0 \re
\left(
\begin{matrix}
v_{\text{in},y} v_{\text{in},z}^*\\
v_{\text{in},x} v_{\text{in},z}^*
\end{matrix}
\right)\\
&+
\frac{\pi {\epsilon}^6}{24 k^3 }   (f_{11} - 2 f_{10} )^2
\rho_0
\im\left(
\begin{matrix}
-v_{\text{in},y} v_{\text{in},z}^*\\
v_{\text{in},x} v_{\text{in},z}^*
\end{matrix}
\right).
\label{tau2}
\end{align}
We  note that
the axial projection of the linear momentum flux is
$({\bf P}\cdot \vec{e}_z)_i=
P_{i,z}=
 (\rho_0/2)\re[v_{\text{in},i}v_{\text{in},z}^*]$, which allows us 
to re-write the $\bigO(\epsilon^3)$ term in the right-hand side of Eq.~\eqref{tau2},
\begin{equation}
\frac{\rho_0}{2} 
\re
\left(
\begin{matrix}
v_{\text{in},y}v_{\text{in},z}^*\\
-v_{\text{in},x}v_{\text{in},z}^*
\end{matrix}
\right)
= -
\left(
\begin{matrix}
[\vec{e}_z \times ( {\bf P} \cdot \vec{e}_z)]_x\\
[\vec{e}_z \times ( {\bf P} \cdot \vec{e}_z)]_y
\end{matrix}
\right).
\label{axialP}
\end{equation}
We turn to the acoustic spin term.
The $\bigO (\epsilon^6)$ term of  Eq.~\eqref{tau2} can be written as
\begin{align}
\nonumber
\rho_0\im\left(
\begin{matrix}
-v_{\text{in},y} v_{\text{in},z}^*\\
v_{\text{in},x} v_{\text{in},z}^*
\end{matrix}
\right) &= \frac{\rho_0}{2}\im 
\left(
\begin{matrix}
(\vec{v}_\text{in}^*\times \vec{v}_\text{in})_x\\
(\vec{v}_\text{in}^*\times \vec{v}_\text{in})_y
\end{matrix}
\right) = \omega 
\left(
\begin{matrix}
S_x\\
S_y
\end{matrix}
\right)\\
&=\omega \vec{S}_\perp,
\label{ST}
\end{align}
with $\vec{S}_\perp$ being the transverse spin density.
By combining Eqs.~\eqref{axialP},
\eqref{ST} into Eq.~\eqref{tau2}, we arrive at
\begin{equation}
\vec{\tau}^\text{rad}
= -\pi a^3 \biggl[   \chi 
\left[\vec{e}_z\times {\bf P}({0})\cdot \vec{e}_z\right] -
\frac{\epsilon^3}{24} \chi^2   \omega    \vec{S}_\perp({0})\biggr],	
\end{equation}
where  $\chi=f_{11}- 2 f_{10} $.

\section{Basic properties of dyads}
\label{app:dyads}
Let $\vec{a}=a_i \vec{e}_i$, $\vec{b}=b_i \vec{e}_i$, and $\vec{c}=c_i \vec{e}_i$ be vectors with $a_i,b_i,c_i\in \mathbb{C}$.
The dyad ${\bf D}$ is defined as the product 
\begin{subequations}
	\begin{align}
		{\bf D} &= \vec{a}\vec{b},\\
		D_{ij} &= a_i b_j, \quad i,j\in \{x,y,z\},
	\end{align}
\end{subequations}
which is a second-rank tensor.
Note that $\vec{ab}\neq\vec{ba}$.
Dyads follow the distributive rule $\vec{a}(\vec{b}+\vec{c})=\vec{a}\vec{b}+\vec{a}\vec{c}$.
The unit dyad is 
\begin{equation}
\label{unitTensor}
{\bf I} = \vec{e}_x\vec{e}_x  + \vec{e}_y\vec{e}_y  + \vec{e}_z\vec{e}_z =\vec{e}_i\vec{e}_i.
\end{equation}
The pre- and post-dot product are defined, respectively, by 
\begin{subequations}
	\begin{align}
		\vec{c}\cdot \vec{ab}&=(\vec{c}\cdot\vec{a})\vec{b} = c_i a_i b_j \vec{e}_j, \\
		\vec{ab}\cdot\vec{c}&=\vec{a}(\vec{b}\cdot \vec{c}) = b_ic_i a_j \vec{e}_j.
	\end{align}
\end{subequations}
We also have
\begin{equation}
\vec{a} \cdot  {\bf I} = {\bf I} \cdot \vec{a} = \vec{a}.
\end{equation}


\begin{thebibliography}{64}%
	\makeatletter
	\providecommand \@ifxundefined [1]{%
		\@ifx{#1\undefined}
	}%
	\providecommand \@ifnum [1]{%
		\ifnum #1\expandafter \@firstoftwo
		\else \expandafter \@secondoftwo
		\fi
	}%
	\providecommand \@ifx [1]{%
		\ifx #1\expandafter \@firstoftwo
		\else \expandafter \@secondoftwo
		\fi
	}%
	\providecommand \natexlab [1]{#1}%
	\providecommand \enquote  [1]{``#1''}%
	\providecommand \bibnamefont  [1]{#1}%
	\providecommand \bibfnamefont [1]{#1}%
	\providecommand \citenamefont [1]{#1}%
	\providecommand \href@noop [0]{\@secondoftwo}%
	\providecommand \href [0]{\begingroup \@sanitize@url \@href}%
	\providecommand \@href[1]{\@@startlink{#1}\@@href}%
	\providecommand \@@href[1]{\endgroup#1\@@endlink}%
	\providecommand \@sanitize@url [0]{\catcode `\\12\catcode `\$12\catcode
		`\&12\catcode `\#12\catcode `\^12\catcode `\_12\catcode `\%12\relax}%
	\providecommand \@@startlink[1]{}%
	\providecommand \@@endlink[0]{}%
	\providecommand \url  [0]{\begingroup\@sanitize@url \@url }%
	\providecommand \@url [1]{\endgroup\@href {#1}{\urlprefix }}%
	\providecommand \urlprefix  [0]{URL }%
	\providecommand \Eprint [0]{\href }%
	\providecommand \doibase [0]{https://doi.org/}%
	\providecommand \selectlanguage [0]{\@gobble}%
	\providecommand \bibinfo  [0]{\@secondoftwo}%
	\providecommand \bibfield  [0]{\@secondoftwo}%
	\providecommand \translation [1]{[#1]}%
	\providecommand \BibitemOpen [0]{}%
	\providecommand \bibitemStop [0]{}%
	\providecommand \bibitemNoStop [0]{.\EOS\space}%
	\providecommand \EOS [0]{\spacefactor3000\relax}%
	\providecommand \BibitemShut  [1]{\csname bibitem#1\endcsname}%
	\let\auto@bib@innerbib\@empty
	\bibitem [{\citenamefont {Ozcelik}\ \emph {et~al.}(2018)\citenamefont
		{Ozcelik}, \citenamefont {Rufo}, \citenamefont {Guo}, \citenamefont {Gu},
		\citenamefont {Li}, \citenamefont {Lata},\ and\ \citenamefont
		{Huang}}]{Ozcelik2018}%
	\BibitemOpen
	\bibfield  {author} {\bibinfo {author} {\bibfnamefont {A.}~\bibnamefont
			{Ozcelik}}, \bibinfo {author} {\bibfnamefont {J.}~\bibnamefont {Rufo}},
		\bibinfo {author} {\bibfnamefont {F.}~\bibnamefont {Guo}}, \bibinfo {author}
		{\bibfnamefont {Y.}~\bibnamefont {Gu}}, \bibinfo {author} {\bibfnamefont
			{P.}~\bibnamefont {Li}}, \bibinfo {author} {\bibfnamefont {J.}~\bibnamefont
			{Lata}},\ and\ \bibinfo {author} {\bibfnamefont {T.~J.}\ \bibnamefont
			{Huang}},\ }\bibfield  {title} {\bibinfo {title} {Acoustic tweezers for the
			life sciences},\ }\href@noop {} {\bibfield  {journal} {\bibinfo  {journal}
			{Nat. Methods}\ }\textbf {\bibinfo {volume} {15}},\ \bibinfo {pages} {1021}
		(\bibinfo {year} {2018})}\BibitemShut {NoStop}%
	\bibitem [{\citenamefont {Wu}\ \emph {et~al.}(2019)\citenamefont {Wu},
		\citenamefont {Ozcelik}, \citenamefont {Rufo}, \citenamefont {Wang},
		\citenamefont {Fang},\ and\ \citenamefont {Huang}}]{Wu2019}%
	\BibitemOpen
	\bibfield  {author} {\bibinfo {author} {\bibfnamefont {M.}~\bibnamefont
			{Wu}}, \bibinfo {author} {\bibfnamefont {A.}~\bibnamefont {Ozcelik}},
		\bibinfo {author} {\bibfnamefont {J.}~\bibnamefont {Rufo}}, \bibinfo {author}
		{\bibfnamefont {Z.}~\bibnamefont {Wang}}, \bibinfo {author} {\bibfnamefont
			{R.}~\bibnamefont {Fang}},\ and\ \bibinfo {author} {\bibfnamefont {T.~J.}\
			\bibnamefont {Huang}},\ }\bibfield  {title} {\bibinfo {title} {Acoustofluidic
			separation of cells and particles},\ }\href@noop {} {\bibfield  {journal}
		{\bibinfo  {journal} {Microsyst. Nanoeng.}\ }\textbf {\bibinfo {volume}
			{5}},\ \bibinfo {pages} {32} (\bibinfo {year} {2019})}\BibitemShut {NoStop}%
	\bibitem [{\citenamefont {Li}\ \emph {et~al.}(2015)\citenamefont {Li},
		\citenamefont {Mao}, \citenamefont {Peng}, \citenamefont {Zhou},
		\citenamefont {Chen}, \citenamefont {Huang}, \citenamefont {Truica},
		\citenamefont {Drabick}, \citenamefont {El-Deiry}, \citenamefont {Dao},
		\citenamefont {Suresh},\ and\ \citenamefont {Huang}}]{Li2015}%
	\BibitemOpen
	\bibfield  {author} {\bibinfo {author} {\bibfnamefont {P.}~\bibnamefont
			{Li}}, \bibinfo {author} {\bibfnamefont {Z.}~\bibnamefont {Mao}}, \bibinfo
		{author} {\bibfnamefont {Z.}~\bibnamefont {Peng}}, \bibinfo {author}
		{\bibfnamefont {L.}~\bibnamefont {Zhou}}, \bibinfo {author} {\bibfnamefont
			{Y.}~\bibnamefont {Chen}}, \bibinfo {author} {\bibfnamefont {P.-H.}\
			\bibnamefont {Huang}}, \bibinfo {author} {\bibfnamefont {C.~I.}\ \bibnamefont
			{Truica}}, \bibinfo {author} {\bibfnamefont {J.~J.}\ \bibnamefont {Drabick}},
		\bibinfo {author} {\bibfnamefont {W.~S.}\ \bibnamefont {El-Deiry}}, \bibinfo
		{author} {\bibfnamefont {M.}~\bibnamefont {Dao}}, \bibinfo {author}
		{\bibfnamefont {S.}~\bibnamefont {Suresh}},\ and\ \bibinfo {author}
		{\bibfnamefont {T.~J.}\ \bibnamefont {Huang}},\ }\bibfield  {title} {\bibinfo
		{title} {Acoustic separation of circulating tumor cells},\ }\href@noop {}
	{\bibfield  {journal} {\bibinfo  {journal} {Proc. Natl. Acad. Sci. USA}\
		}\textbf {\bibinfo {volume} {112}},\ \bibinfo {pages} {4970} (\bibinfo {year}
		{2015})}\BibitemShut {NoStop}%
	\bibitem [{\citenamefont {Collins}\ \emph {et~al.}(2015)\citenamefont
		{Collins}, \citenamefont {Morahan}, \citenamefont {Garcia-Bustos},
		\citenamefont {Plebanski},\ and\ \citenamefont {Neild}}]{Collins2015}%
	\BibitemOpen
	\bibfield  {author} {\bibinfo {author} {\bibfnamefont {D.~J.}\ \bibnamefont
			{Collins}}, \bibinfo {author} {\bibfnamefont {B.}~\bibnamefont {Morahan}},
		\bibinfo {author} {\bibfnamefont {J.}~\bibnamefont {Garcia-Bustos}}, \bibinfo
		{author} {\bibfnamefont {C.~D.~M.}\ \bibnamefont {Plebanski}},\ and\ \bibinfo
		{author} {\bibfnamefont {A.}~\bibnamefont {Neild}},\ }\bibfield  {title}
	{\bibinfo {title} {Two-dimensional single-cell patterning with one cell per
			well driven by surface acoustic waves},\ }\href@noop {} {\bibfield  {journal}
		{\bibinfo  {journal} {Nat. Commun.}\ }\textbf {\bibinfo {volume} {6}},\
		\bibinfo {pages} {8686} (\bibinfo {year} {2015})}\BibitemShut {NoStop}%
	\bibitem [{\citenamefont {Silva}\ \emph
		{et~al.}(2019{\natexlab{a}})\citenamefont {Silva}, \citenamefont {Lopes},
		\citenamefont {Leao-Neto}, \citenamefont {Nichols},\ and\ \citenamefont
		{Drinkwater}}]{Silva2019}%
	\BibitemOpen
	\bibfield  {author} {\bibinfo {author} {\bibfnamefont {G.~T.}\ \bibnamefont
			{Silva}}, \bibinfo {author} {\bibfnamefont {J.~H.}\ \bibnamefont {Lopes}},
		\bibinfo {author} {\bibfnamefont {J.~P.}\ \bibnamefont {Leao-Neto}}, \bibinfo
		{author} {\bibfnamefont {M.~K.}\ \bibnamefont {Nichols}},\ and\ \bibinfo
		{author} {\bibfnamefont {B.~W.}\ \bibnamefont {Drinkwater}},\ }\bibfield
	{title} {\bibinfo {title} {Particle patterning by ultrasonic standing waves
			in a rectangular cavity},\ }\href@noop {} {\bibfield  {journal} {\bibinfo
			{journal} {Phys. Rev. Applied}\ }\textbf {\bibinfo {volume} {11}},\ \bibinfo
		{pages} {054044} (\bibinfo {year} {2019}{\natexlab{a}})}\BibitemShut
	{NoStop}%
	\bibitem [{\citenamefont {Mishra}\ \emph {et~al.}(2014)\citenamefont {Mishra},
		\citenamefont {Hill},\ and\ \citenamefont {Glynne-Jones}}]{Mishra2014}%
	\BibitemOpen
	\bibfield  {author} {\bibinfo {author} {\bibfnamefont {P.}~\bibnamefont
			{Mishra}}, \bibinfo {author} {\bibfnamefont {M.}~\bibnamefont {Hill}},\ and\
		\bibinfo {author} {\bibfnamefont {P.}~\bibnamefont {Glynne-Jones}},\
	}\bibfield  {title} {\bibinfo {title} {Deformation of red blood cells using
			acoustic radiation forces},\ }\href@noop {} {\bibfield  {journal} {\bibinfo
			{journal} {Biomicrofluidics}\ }\textbf {\bibinfo {volume} {8}},\ \bibinfo
		{pages} {034109} (\bibinfo {year} {2014})}\BibitemShut {NoStop}%
	\bibitem [{\citenamefont {Silva}\ \emph
		{et~al.}(2019{\natexlab{b}})\citenamefont {Silva}, \citenamefont {Tian},
		\citenamefont {Franklin}, \citenamefont {Wang}, \citenamefont {Han},
		\citenamefont {Mann},\ and\ \citenamefont {Drinkwater}}]{Silva2019a}%
	\BibitemOpen
	\bibfield  {author} {\bibinfo {author} {\bibfnamefont {G.~T.}\ \bibnamefont
			{Silva}}, \bibinfo {author} {\bibfnamefont {L.}~\bibnamefont {Tian}},
		\bibinfo {author} {\bibfnamefont {A.}~\bibnamefont {Franklin}}, \bibinfo
		{author} {\bibfnamefont {X.}~\bibnamefont {Wang}}, \bibinfo {author}
		{\bibfnamefont {X.}~\bibnamefont {Han}}, \bibinfo {author} {\bibfnamefont
			{S.}~\bibnamefont {Mann}},\ and\ \bibinfo {author} {\bibfnamefont {B.~W.}\
			\bibnamefont {Drinkwater}},\ }\bibfield  {title} {\bibinfo {title} {Acoustic
			deformation for the extraction of mechanical properties of lipid vesicle
			populations},\ }\href@noop {} {\bibfield  {journal} {\bibinfo  {journal}
			{Phys. Rev. E}\ }\textbf {\bibinfo {volume} {99}},\ \bibinfo {pages} {063002}
		(\bibinfo {year} {2019}{\natexlab{b}})}\BibitemShut {NoStop}%
	\bibitem [{\citenamefont {Baudoin}\ \emph {et~al.}(2019)\citenamefont
		{Baudoin}, \citenamefont {Gerbedoen}, \citenamefont {Riaud}, \citenamefont
		{Matar}, \citenamefont {Smagin},\ and\ \citenamefont {Thomas}}]{Baudoin2019}%
	\BibitemOpen
	\bibfield  {author} {\bibinfo {author} {\bibfnamefont {M.}~\bibnamefont
			{Baudoin}}, \bibinfo {author} {\bibfnamefont {J.-C.}\ \bibnamefont
			{Gerbedoen}}, \bibinfo {author} {\bibfnamefont {A.}~\bibnamefont {Riaud}},
		\bibinfo {author} {\bibfnamefont {O.~B.}\ \bibnamefont {Matar}}, \bibinfo
		{author} {\bibfnamefont {N.}~\bibnamefont {Smagin}},\ and\ \bibinfo {author}
		{\bibfnamefont {J.-L.}\ \bibnamefont {Thomas}},\ }\bibfield  {title}
	{\bibinfo {title} {Folding a focalized acoustical vortex on a flat
			holographic transducer: Miniaturized selective acoustical tweezers},\
	}\href@noop {} {\bibfield  {journal} {\bibinfo  {journal} {Sci. Adv.}\
		}\textbf {\bibinfo {volume} {5}},\ \bibinfo {pages} {eaav1967} (\bibinfo
		{year} {2019})}\BibitemShut {NoStop}%
	\bibitem [{\citenamefont {Baudoin}\ and\ \citenamefont
		{Thomas}(2019)}]{Baudoin2019b}%
	\BibitemOpen
	\bibfield  {author} {\bibinfo {author} {\bibfnamefont {M.}~\bibnamefont
			{Baudoin}}\ and\ \bibinfo {author} {\bibfnamefont {J.-L.}\ \bibnamefont
			{Thomas}},\ }\bibfield  {title} {\bibinfo {title} {Acoustic tweezers for
			particle and fluid micromanipulation},\ }\href@noop {} {\bibfield  {journal}
		{\bibinfo  {journal} {Annu. Rev. Fluid Mech.}\ }\textbf {\bibinfo {volume}
			{52}},\ \bibinfo {pages} {205} (\bibinfo {year} {2019})}\BibitemShut
	{NoStop}%
	\bibitem [{\citenamefont {Schwarz}\ \emph {et~al.}(2012)\citenamefont
		{Schwarz}, \citenamefont {Petit-Pierre},\ and\ \citenamefont
		{Dual}}]{Schwarz2012}%
	\BibitemOpen
	\bibfield  {author} {\bibinfo {author} {\bibfnamefont {T.}~\bibnamefont
			{Schwarz}}, \bibinfo {author} {\bibfnamefont {G.}~\bibnamefont
			{Petit-Pierre}},\ and\ \bibinfo {author} {\bibfnamefont {J.}~\bibnamefont
			{Dual}},\ }\bibfield  {title} {\bibinfo {title} {Rotation of non-spherical
			micro-particles by amplitude modulation of superimposed orthogonal ultrasonic
			modes},\ }\href@noop {} {\bibfield  {journal} {\bibinfo  {journal} {J.
				Acoust. Soc. Am.}\ }\textbf {\bibinfo {volume} {133}},\ \bibinfo {pages}
		{1260} (\bibinfo {year} {2012})}\BibitemShut {NoStop}%
	\bibitem [{\citenamefont {Ai}\ \emph {et~al.}(2013)\citenamefont {Ai},
		\citenamefont {Sanders},\ and\ \citenamefont {Marrone}}]{Ai2013}%
	\BibitemOpen
	\bibfield  {author} {\bibinfo {author} {\bibfnamefont {Y.}~\bibnamefont
			{Ai}}, \bibinfo {author} {\bibfnamefont {C.~K.}\ \bibnamefont {Sanders}},\
		and\ \bibinfo {author} {\bibfnamefont {B.~L.}\ \bibnamefont {Marrone}},\
	}\bibfield  {title} {\bibinfo {title} {Separation of escherichia coli
			bacteria from peripheral blood mononuclear cells using standing surface
			acoustic waves},\ }\href@noop {} {\bibfield  {journal} {\bibinfo  {journal}
			{Anal. Chem.}\ }\textbf {\bibinfo {volume} {85}},\ \bibinfo {pages} {9126}
		(\bibinfo {year} {2013})}\BibitemShut {NoStop}%
	\bibitem [{\citenamefont {Jakobsson}\ \emph {et~al.}(2014)\citenamefont
		{Jakobsson}, \citenamefont {Antfolk},\ and\ \citenamefont
		{Laurell}}]{Jakobsson2014}%
	\BibitemOpen
	\bibfield  {author} {\bibinfo {author} {\bibfnamefont {O.}~\bibnamefont
			{Jakobsson}}, \bibinfo {author} {\bibfnamefont {M.}~\bibnamefont {Antfolk}},\
		and\ \bibinfo {author} {\bibfnamefont {T.}~\bibnamefont {Laurell}},\
	}\bibfield  {title} {\bibinfo {title} {Continuous flow two-dimensional
			acoustic orientation of nonspherical cells},\ }\href@noop {} {\bibfield
		{journal} {\bibinfo  {journal} {Anal. Chem.}\ }\textbf {\bibinfo {volume}
			{86}},\ \bibinfo {pages} {6111} (\bibinfo {year} {2014})}\BibitemShut
	{NoStop}%
	\bibitem [{\citenamefont {Schwarz}\ \emph {et~al.}(2015)\citenamefont
		{Schwarz}, \citenamefont {Hahn}, \citenamefont {Petit-Pierre},\ and\
		\citenamefont {Dual}}]{Schwarz2015}%
	\BibitemOpen
	\bibfield  {author} {\bibinfo {author} {\bibfnamefont {T.}~\bibnamefont
			{Schwarz}}, \bibinfo {author} {\bibfnamefont {P.}~\bibnamefont {Hahn}},
		\bibinfo {author} {\bibfnamefont {G.}~\bibnamefont {Petit-Pierre}},\ and\
		\bibinfo {author} {\bibfnamefont {J.}~\bibnamefont {Dual}},\ }\bibfield
	{title} {\bibinfo {title} {Rotation of fibers and other non-spherical
			particles by the acoustic radiation torque},\ }\href@noop {} {\bibfield
		{journal} {\bibinfo  {journal} {Microfluid Nanofluid}\ }\textbf {\bibinfo
			{volume} {18}},\ \bibinfo {pages} {65} (\bibinfo {year} {2015})}\BibitemShut
	{NoStop}%
	\bibitem [{\citenamefont {Garbin}\ \emph {et~al.}(2015)\citenamefont {Garbin},
		\citenamefont {Leibacher}, \citenamefont {Hahn}, \citenamefont {Ferrand},
		\citenamefont {Studart},\ and\ \citenamefont {Dual}}]{Garbin2015}%
	\BibitemOpen
	\bibfield  {author} {\bibinfo {author} {\bibfnamefont {A.}~\bibnamefont
			{Garbin}}, \bibinfo {author} {\bibfnamefont {I.}~\bibnamefont {Leibacher}},
		\bibinfo {author} {\bibfnamefont {P.}~\bibnamefont {Hahn}}, \bibinfo {author}
		{\bibfnamefont {H.~L.}\ \bibnamefont {Ferrand}}, \bibinfo {author}
		{\bibfnamefont {A.}~\bibnamefont {Studart}},\ and\ \bibinfo {author}
		{\bibfnamefont {J.}~\bibnamefont {Dual}},\ }\bibfield  {title} {\bibinfo
		{title} {Acoustophoresis of disk-shaped microparticles: A numerical and
			experimental study of acoustic radiation forces and torques},\ }\href@noop {}
	{\bibfield  {journal} {\bibinfo  {journal} {J. Acoust. Soc. Am.}\ }\textbf
		{\bibinfo {volume} {138}},\ \bibinfo {pages} {2759} (\bibinfo {year}
		{2015})}\BibitemShut {NoStop}%
	\bibitem [{\citenamefont {Foresti}\ and\ \citenamefont
		{Poulikakos}(2014)}]{Foresti2013}%
	\BibitemOpen
	\bibfield  {author} {\bibinfo {author} {\bibfnamefont {D.}~\bibnamefont
			{Foresti}}\ and\ \bibinfo {author} {\bibfnamefont {D.}~\bibnamefont
			{Poulikakos}},\ }\bibfield  {title} {\bibinfo {title} {{Acoustophoretic
				contactless elevation, orbital transport and spinning of matter in air}},\
	}\href {https://doi.org/10.1103/PhysRevLett.112.024301} {\bibfield  {journal}
		{\bibinfo  {journal} {Phys. Rev. Lett.}\ }\textbf {\bibinfo {volume} {112}},\
		\bibinfo {pages} {024301} (\bibinfo {year} {2014})}\BibitemShut {NoStop}%
	\bibitem [{\citenamefont {Marzo}\ \emph {et~al.}(2015)\citenamefont {Marzo},
		\citenamefont {Seah}, \citenamefont {Drinkwater}, \citenamefont {Sahoo},
		\citenamefont {Long},\ and\ \citenamefont {Subramanian}}]{Marzo2015}%
	\BibitemOpen
	\bibfield  {author} {\bibinfo {author} {\bibfnamefont {A.}~\bibnamefont
			{Marzo}}, \bibinfo {author} {\bibfnamefont {S.~A.}\ \bibnamefont {Seah}},
		\bibinfo {author} {\bibfnamefont {B.~W.}\ \bibnamefont {Drinkwater}},
		\bibinfo {author} {\bibfnamefont {D.~R.}\ \bibnamefont {Sahoo}}, \bibinfo
		{author} {\bibfnamefont {B.}~\bibnamefont {Long}},\ and\ \bibinfo {author}
		{\bibfnamefont {S.}~\bibnamefont {Subramanian}},\ }\bibfield  {title}
	{\bibinfo {title} {{Holographic acoustic elements for manipulation of
				levitated objects}},\ }\href@noop {} {\bibfield  {journal} {\bibinfo
			{journal} {Nat. Commun.}\ }\textbf {\bibinfo {volume} {6}},\ \bibinfo {pages}
		{8661} (\bibinfo {year} {2015})}\BibitemShut {NoStop}%
	\bibitem [{\citenamefont {Hong}\ \emph {et~al.}(2017)\citenamefont {Hong},
		\citenamefont {Yin}, \citenamefont {Zhai}, \citenamefont {Yan}, \citenamefont
		{Wang}, \citenamefont {Zhang},\ and\ \citenamefont {Drinkwater}}]{Hong2017}%
	\BibitemOpen
	\bibfield  {author} {\bibinfo {author} {\bibfnamefont {Z.~Y.}\ \bibnamefont
			{Hong}}, \bibinfo {author} {\bibfnamefont {J.~F.}\ \bibnamefont {Yin}},
		\bibinfo {author} {\bibfnamefont {W.}~\bibnamefont {Zhai}}, \bibinfo {author}
		{\bibfnamefont {N.}~\bibnamefont {Yan}}, \bibinfo {author} {\bibfnamefont
			{W.~L.}\ \bibnamefont {Wang}}, \bibinfo {author} {\bibfnamefont
			{J.}~\bibnamefont {Zhang}},\ and\ \bibinfo {author} {\bibfnamefont {B.~W.}\
			\bibnamefont {Drinkwater}},\ }\bibfield  {title} {\bibinfo {title} {Dynamics
			of levitated objects in acoustic vortex fields},\ }\href@noop {} {\bibfield
		{journal} {\bibinfo  {journal} {Sci. Rep.}\ }\textbf {\bibinfo {volume}
			{7}},\ \bibinfo {pages} {7093} (\bibinfo {year} {2017})}\BibitemShut
	{NoStop}%
	\bibitem [{\citenamefont {Wang}\ \emph {et~al.}(2012)\citenamefont {Wang},
		\citenamefont {Castro}, \citenamefont {Hoyos},\ and\ \citenamefont
		{Mallouk}}]{Wang2012}%
	\BibitemOpen
	\bibfield  {author} {\bibinfo {author} {\bibfnamefont {W.}~\bibnamefont
			{Wang}}, \bibinfo {author} {\bibfnamefont {L.~A.}\ \bibnamefont {Castro}},
		\bibinfo {author} {\bibfnamefont {M.}~\bibnamefont {Hoyos}},\ and\ \bibinfo
		{author} {\bibfnamefont {T.~E.}\ \bibnamefont {Mallouk}},\ }\bibfield
	{title} {\bibinfo {title} {Autonomous motion of metallic microrods propelled
			by ultrasound},\ }\href@noop {} {\bibfield  {journal} {\bibinfo  {journal}
			{ACS Nano}\ }\textbf {\bibinfo {volume} {67}},\ \bibinfo {pages} {6122}
		(\bibinfo {year} {2012})}\BibitemShut {NoStop}%
	\bibitem [{\citenamefont {Ren}\ \emph {et~al.}(2019)\citenamefont {Ren},
		\citenamefont {Nama}, \citenamefont {McNeill}, \citenamefont {Soto},
		\citenamefont {Yan}, \citenamefont {Liu}, \citenamefont {Wang}, \citenamefont
		{Wang},\ and\ \citenamefont {Mallouk}}]{Ren2019}%
	\BibitemOpen
	\bibfield  {author} {\bibinfo {author} {\bibfnamefont {L.}~\bibnamefont
			{Ren}}, \bibinfo {author} {\bibfnamefont {N.}~\bibnamefont {Nama}}, \bibinfo
		{author} {\bibfnamefont {J.~M.}\ \bibnamefont {McNeill}}, \bibinfo {author}
		{\bibfnamefont {F.}~\bibnamefont {Soto}}, \bibinfo {author} {\bibfnamefont
			{Z.}~\bibnamefont {Yan}}, \bibinfo {author} {\bibfnamefont {W.}~\bibnamefont
			{Liu}}, \bibinfo {author} {\bibfnamefont {W.}~\bibnamefont {Wang}}, \bibinfo
		{author} {\bibfnamefont {J.}~\bibnamefont {Wang}},\ and\ \bibinfo {author}
		{\bibfnamefont {T.~E.}\ \bibnamefont {Mallouk}},\ }\bibfield  {title}
	{\bibinfo {title} {3d steerable, acoustically powered microswimmers for
			single-particle manipulation},\ }\href@noop {} {\bibfield  {journal}
		{\bibinfo  {journal} {Sci. Adv.}\ }\textbf {\bibinfo {volume} {5}},\ \bibinfo
		{pages} {eaax3084} (\bibinfo {year} {2019})}\BibitemShut {NoStop}%
	\bibitem [{\citenamefont {Glynne-Jones}\ \emph {et~al.}(2013)\citenamefont
		{Glynne-Jones}, \citenamefont {Mishra}, \citenamefont {Boltryk},\ and\
		\citenamefont {Hill}}]{Glynne-Jones2013}%
	\BibitemOpen
	\bibfield  {author} {\bibinfo {author} {\bibfnamefont {P.}~\bibnamefont
			{Glynne-Jones}}, \bibinfo {author} {\bibfnamefont {P.~P.}\ \bibnamefont
			{Mishra}}, \bibinfo {author} {\bibfnamefont {R.~J.}\ \bibnamefont
			{Boltryk}},\ and\ \bibinfo {author} {\bibfnamefont {M.}~\bibnamefont
			{Hill}},\ }\bibfield  {title} {\bibinfo {title} {Efficient finite element
			modeling of radiation forces on elastic particles of arbitrary size and
			geometry},\ }\href@noop {} {\bibfield  {journal} {\bibinfo  {journal} {J.
				Acoust. Soc. Am.}\ }\textbf {\bibinfo {volume} {133}},\ \bibinfo {pages}
		{1885} (\bibinfo {year} {2013})}\BibitemShut {NoStop}%
	\bibitem [{\citenamefont {Hahn}\ \emph {et~al.}(2015)\citenamefont {Hahn},
		\citenamefont {Leibacher}, \citenamefont {Baasch},\ and\ \citenamefont
		{Dual}}]{Hahn2015}%
	\BibitemOpen
	\bibfield  {author} {\bibinfo {author} {\bibfnamefont {P.}~\bibnamefont
			{Hahn}}, \bibinfo {author} {\bibfnamefont {I.}~\bibnamefont {Leibacher}},
		\bibinfo {author} {\bibfnamefont {T.}~\bibnamefont {Baasch}},\ and\ \bibinfo
		{author} {\bibfnamefont {J.}~\bibnamefont {Dual}},\ }\bibfield  {title}
	{\bibinfo {title} {Numerical simulation of acoustofluidic manipulation by
			radiation forces and acoustic streaming for complex particles},\ }\href@noop
	{} {\bibfield  {journal} {\bibinfo  {journal} {Lab. Chip.}\ }\textbf
		{\bibinfo {volume} {15}},\ \bibinfo {pages} {4302} (\bibinfo {year}
		{2015})}\BibitemShut {NoStop}%
	\bibitem [{\citenamefont {{Greve}}\ \emph {et~al.}(2018)\citenamefont
		{{Greve}}, \citenamefont {{Dauson}},\ and\ \citenamefont
		{{Oppenheim}}}]{Greve2018}%
	\BibitemOpen
	\bibfield  {author} {\bibinfo {author} {\bibfnamefont {D.~W.}\ \bibnamefont
			{{Greve}}}, \bibinfo {author} {\bibfnamefont {E.~R.}\ \bibnamefont
			{{Dauson}}},\ and\ \bibinfo {author} {\bibfnamefont {I.~J.}\ \bibnamefont
			{{Oppenheim}}},\ }\bibfield  {title} {\bibinfo {title} {Forces and torques on
			rods in an ultrasonic standing wave},\ }in\ \href@noop {} {\emph {\bibinfo
			{booktitle} {2018 IEEE International Ultrasonics Symposium (IUS)}}}\
	(\bibinfo {year} {2018})\ pp.\ \bibinfo {pages} {1--4}\BibitemShut {NoStop}%
	\bibitem [{\citenamefont {Wijaya}\ and\ \citenamefont
		{Lim}(2015)}]{Wijaya2015}%
	\BibitemOpen
	\bibfield  {author} {\bibinfo {author} {\bibfnamefont {F.~B.}\ \bibnamefont
			{Wijaya}}\ and\ \bibinfo {author} {\bibfnamefont {K.-M.}\ \bibnamefont
			{Lim}},\ }\bibfield  {title} {\bibinfo {title} {Numerical calculation of
			acoustic radiation force and torque acting on rigid non-spherical
			particles},\ }\href@noop {} {\bibfield  {journal} {\bibinfo  {journal} {Acta
				Acust. united Ac.}\ }\textbf {\bibinfo {volume} {101}},\ \bibinfo {pages}
		{531} (\bibinfo {year} {2015})}\BibitemShut {NoStop}%
	\bibitem [{\citenamefont {Jerome}\ \emph {et~al.}(2019)\citenamefont {Jerome},
		\citenamefont {Ilinskii}, \citenamefont {Zabolotskaya},\ and\ \citenamefont
		{Hamilton}}]{Jerome2019}%
	\BibitemOpen
	\bibfield  {author} {\bibinfo {author} {\bibfnamefont {T.~S.}\ \bibnamefont
			{Jerome}}, \bibinfo {author} {\bibfnamefont {Y.~A.}\ \bibnamefont
			{Ilinskii}}, \bibinfo {author} {\bibfnamefont {E.~A.}\ \bibnamefont
			{Zabolotskaya}},\ and\ \bibinfo {author} {\bibfnamefont {M.~F.}\ \bibnamefont
			{Hamilton}},\ }\bibfield  {title} {\bibinfo {title} {Born approximation of
			acoustic radiation force and torque on soft objects of arbitrary shape},\
	}\href@noop {} {\bibfield  {journal} {\bibinfo  {journal} {J. Acoust. Soc.
				Am.}\ }\textbf {\bibinfo {volume} {145}},\ \bibinfo {pages} {36} (\bibinfo
		{year} {2019})}\BibitemShut {NoStop}%
	\bibitem [{\citenamefont {Mitri}(2015)}]{Mitri2015}%
	\BibitemOpen
	\bibfield  {author} {\bibinfo {author} {\bibfnamefont {F.~G.}\ \bibnamefont
			{Mitri}},\ }\bibfield  {title} {\bibinfo {title} {Acoustic radiation force on
			oblate and prolate spheroids in {B}essel beams},\ }\href@noop {} {\bibfield
		{journal} {\bibinfo  {journal} {Wave Motion}\ }\textbf {\bibinfo {volume}
			{57}},\ \bibinfo {pages} {231} (\bibinfo {year} {2015})}\BibitemShut
	{NoStop}%
	\bibitem [{\citenamefont {Mitri}(2016)}]{Mitri2016a}%
	\BibitemOpen
	\bibfield  {author} {\bibinfo {author} {\bibfnamefont {F.~G.}\ \bibnamefont
			{Mitri}},\ }\bibfield  {title} {\bibinfo {title} {Radiation forces and torque
			on a rigid elliptical cylinder in acoustical plane progressive and
			(quasi)standing waves with arbitrary incidence},\ }\href@noop {} {\bibfield
		{journal} {\bibinfo  {journal} {Phys. Fluid}\ }\textbf {\bibinfo {volume}
			{28}},\ \bibinfo {pages} {077104} (\bibinfo {year} {2016})}\BibitemShut
	{NoStop}%
	\bibitem [{\citenamefont {Gong}\ \emph
		{et~al.}(2019{\natexlab{a}})\citenamefont {Gong}, \citenamefont {Marston},\
		and\ \citenamefont {Li}}]{Gong2019a}%
	\BibitemOpen
	\bibfield  {author} {\bibinfo {author} {\bibfnamefont {Z.}~\bibnamefont
			{Gong}}, \bibinfo {author} {\bibfnamefont {P.~L.}\ \bibnamefont {Marston}},\
		and\ \bibinfo {author} {\bibfnamefont {W.}~\bibnamefont {Li}},\ }\bibfield
	{title} {\bibinfo {title} {{$T$-matrix evaluation of three-dimensional
				acoustic radiation forces on nonspherical objects in Bessel beams with
				arbitrary order and location}},\ }\href@noop {} {\bibfield  {journal}
		{\bibinfo  {journal} {Phys. Rev. E}\ }\textbf {\bibinfo {volume} {99}},\
		\bibinfo {pages} {063004} (\bibinfo {year} {2019}{\natexlab{a}})}\BibitemShut
	{NoStop}%
	\bibitem [{\citenamefont {Kotani}(1933)}]{Kotani1933}%
	\BibitemOpen
	\bibfield  {author} {\bibinfo {author} {\bibfnamefont {M.}~\bibnamefont
			{Kotani}},\ }\bibfield  {title} {\bibinfo {title} {An acoustical problem
			relating to the theory of {Rayleigh} disc},\ }\href@noop {} {\bibfield
		{journal} {\bibinfo  {journal} {Proc. Phys. Math. Soc. Japan}\ }\textbf
		{\bibinfo {volume} {15}},\ \bibinfo {pages} {30} (\bibinfo {year}
		{1933})}\BibitemShut {NoStop}%
	\bibitem [{\citenamefont {King}(1935)}]{King1935}%
	\BibitemOpen
	\bibfield  {author} {\bibinfo {author} {\bibfnamefont {L.~V.}\ \bibnamefont
			{King}},\ }\bibfield  {title} {\bibinfo {title} {{On the theory of the
				inertia and diffraction corrections for the Rayleigh disc}},\ }\href@noop {}
	{\bibfield  {journal} {\bibinfo  {journal} {Proc. Royal Soc. A}\ }\textbf
		{\bibinfo {volume} {153}},\ \bibinfo {pages} {17} (\bibinfo {year}
		{1935})}\BibitemShut {NoStop}%
	\bibitem [{\citenamefont {Keller}(1957)}]{Keller1957}%
	\BibitemOpen
	\bibfield  {author} {\bibinfo {author} {\bibfnamefont {J.~B.}\ \bibnamefont
			{Keller}},\ }\bibfield  {title} {\bibinfo {title} {Acoustic torques and
			forces on disks},\ }\href@noop {} {\bibfield  {journal} {\bibinfo  {journal}
			{J. Acoust. Soc. Am.}\ }\textbf {\bibinfo {volume} {29}},\ \bibinfo {pages}
		{1085} (\bibinfo {year} {1957})}\BibitemShut {NoStop}%
	\bibitem [{\citenamefont {Maidanik}(1958)}]{Maidanik1958}%
	\BibitemOpen
	\bibfield  {author} {\bibinfo {author} {\bibfnamefont {G.}~\bibnamefont
			{Maidanik}},\ }\bibfield  {title} {\bibinfo {title} {{Torques due to
				acoustical radiation pressure}},\ }\href {https://doi.org/10.1121/1.1909714}
	{\bibfield  {journal} {\bibinfo  {journal} {J. Acoust. Soc. Am.}\ }\textbf
		{\bibinfo {volume} {30}},\ \bibinfo {pages} {620} (\bibinfo {year}
		{1958})}\BibitemShut {NoStop}%
	\bibitem [{\citenamefont {Marston}(2016)}]{Marston2016}%
	\BibitemOpen
	\bibfield  {author} {\bibinfo {author} {\bibfnamefont {P.~L.}\ \bibnamefont
			{Marston}},\ }\bibfield  {title} {\bibinfo {title} {Comment on “radiation
			forces and torque on a rigid elliptical cylinder in acoustical plane
			progressive and (quasi)standing waves with arbitrary incidence” {[Phys.
				Fluids 28, 077104 (2016)]}},\ }\href@noop {} {\bibfield  {journal} {\bibinfo
			{journal} {Phys. Fluids}\ }\textbf {\bibinfo {volume} {29}},\ \bibinfo
		{pages} {029101} (\bibinfo {year} {2016})}\BibitemShut {NoStop}%
	\bibitem [{\citenamefont {Marston}\ \emph {et~al.}(2006)\citenamefont
		{Marston}, \citenamefont {Wei},\ and\ \citenamefont
		{Thiessen}}]{Marston2006a}%
	\BibitemOpen
	\bibfield  {author} {\bibinfo {author} {\bibfnamefont {P.~L.}\ \bibnamefont
			{Marston}}, \bibinfo {author} {\bibfnamefont {W.}~\bibnamefont {Wei}},\ and\
		\bibinfo {author} {\bibfnamefont {D.~B.}\ \bibnamefont {Thiessen}},\
	}\bibfield  {title} {\bibinfo {title} {Acoustic radiation force on elliptical
			cylinders and spheroidal objects in low frequency standing waves},\ }in\
	\href@noop {} {\emph {\bibinfo {booktitle} {AIP Conf. Proc.}}},\ \bibinfo
	{series and number} {\bibinfo {number} {838}}\ (\bibinfo {year} {2006})\ pp.\
	\bibinfo {pages} {495--499}\BibitemShut {NoStop}%
	\bibitem [{\citenamefont {Silva}\ and\ \citenamefont
		{Drinkwater}(2018)}]{Silva2018}%
	\BibitemOpen
	\bibfield  {author} {\bibinfo {author} {\bibfnamefont {G.~T.}\ \bibnamefont
			{Silva}}\ and\ \bibinfo {author} {\bibfnamefont {B.~W.}\ \bibnamefont
			{Drinkwater}},\ }\bibfield  {title} {\bibinfo {title} {Acoustic radiation
			force exerted on a small spheroidal rigid particle by a beam of arbitrary
			wavefront: Examples of traveling and standing plane waves},\ }\href@noop {}
	{\bibfield  {journal} {\bibinfo  {journal} {J. Acoustic. Soc. Am.}\ }\textbf
		{\bibinfo {volume} {144}},\ \bibinfo {pages} {EL453} (\bibinfo {year}
		{2018})}\BibitemShut {NoStop}%
	\bibitem [{\citenamefont {Fan}\ \emph {et~al.}(2008)\citenamefont {Fan},
		\citenamefont {Mei}, \citenamefont {Yang},\ and\ \citenamefont
		{Chen}}]{Fan2008}%
	\BibitemOpen
	\bibfield  {author} {\bibinfo {author} {\bibfnamefont {Z.}~\bibnamefont
			{Fan}}, \bibinfo {author} {\bibfnamefont {D.}~\bibnamefont {Mei}}, \bibinfo
		{author} {\bibfnamefont {K.}~\bibnamefont {Yang}},\ and\ \bibinfo {author}
		{\bibfnamefont {Z.}~\bibnamefont {Chen}},\ }\bibfield  {title} {\bibinfo
		{title} {Acoustic radiation torque on an irregularly shaped scatterer in an
			arbitrary sound field},\ }\href@noop {} {\bibfield  {journal} {\bibinfo
			{journal} {J. Acoust. Soc. Am.}\ }\textbf {\bibinfo {volume} {124}},\
		\bibinfo {pages} {2727} (\bibinfo {year} {2008})}\BibitemShut {NoStop}%
	\bibitem [{\citenamefont {{Le\~ao}-Neto}\ \emph {et~al.}(2020)\citenamefont
		{{Le\~ao}-Neto}, \citenamefont {Lopes},\ and\ \citenamefont
		{Silva}}]{Leao-Neto2020}%
	\BibitemOpen
	\bibfield  {author} {\bibinfo {author} {\bibfnamefont {J.~P.}\ \bibnamefont
			{{Le\~ao}-Neto}}, \bibinfo {author} {\bibfnamefont {J.~H.}\ \bibnamefont
			{Lopes}},\ and\ \bibinfo {author} {\bibfnamefont {G.~T.}\ \bibnamefont
			{Silva}},\ }\bibfield  {title} {\bibinfo {title} {Acoustic radiation torque
			exerted on a subwavelength spheroidal particle by a traveling and standing
			plane wave},\ }\href@noop {} {\bibfield  {journal} {\bibinfo  {journal} {J.
				Acoust. Soc. Am.}\ }\textbf {\bibinfo {volume} {147}},\ \bibinfo {pages}
		{2177} (\bibinfo {year} {2020})},\ \bibinfo {note} {submitted for
		publication}\BibitemShut {NoStop}%
	\bibitem [{\citenamefont {Lopes}\ \emph {et~al.}(2020)\citenamefont {Lopes},
		\citenamefont {Lima}, \citenamefont {{Le\~ao}-Neto},\ and\ \citenamefont
		{Silva}}]{Lopes2020}%
	\BibitemOpen
	\bibfield  {author} {\bibinfo {author} {\bibfnamefont {J.~H.}\ \bibnamefont
			{Lopes}}, \bibinfo {author} {\bibfnamefont {E.~B.}\ \bibnamefont {Lima}},
		\bibinfo {author} {\bibfnamefont {J.~P.}\ \bibnamefont {{Le\~ao}-Neto}},\
		and\ \bibinfo {author} {\bibfnamefont {G.~T.}\ \bibnamefont {Silva}},\
	}\bibfield  {title} {\bibinfo {title} {Acoustic spin transfer to a
			subwavelength spheroidal particle},\ }\href@noop {} {\bibfield  {journal}
		{\bibinfo  {journal} {Phys. Rev. E}\ }\textbf {\bibinfo {volume} {101}},\
		\bibinfo {pages} {043102} (\bibinfo {year} {2020})}\BibitemShut {NoStop}%
	\bibitem [{\citenamefont {Baresch}\ \emph {et~al.}(2013)\citenamefont
		{Baresch}, \citenamefont {Thomas},\ and\ \citenamefont
		{Marchiano}}]{Baresch2013}%
	\BibitemOpen
	\bibfield  {author} {\bibinfo {author} {\bibfnamefont {D.}~\bibnamefont
			{Baresch}}, \bibinfo {author} {\bibfnamefont {J.-L.}\ \bibnamefont
			{Thomas}},\ and\ \bibinfo {author} {\bibfnamefont {R.}~\bibnamefont
			{Marchiano}},\ }\bibfield  {title} {\bibinfo {title} {{Spherical vortex beams
				of high radial degree for enhanced single-beam tweezers}},\ }\href@noop {}
	{\bibfield  {journal} {\bibinfo  {journal} {J. Appl. Phys.}\ }\textbf
		{\bibinfo {volume} {113}},\ \bibinfo {pages} {184901} (\bibinfo {year}
		{2013})}\BibitemShut {NoStop}%
	\bibitem [{\citenamefont {Mitri}\ and\ \citenamefont
		{Silva}(2014)}]{Mitri2014}%
	\BibitemOpen
	\bibfield  {author} {\bibinfo {author} {\bibfnamefont {F.~G.}\ \bibnamefont
			{Mitri}}\ and\ \bibinfo {author} {\bibfnamefont {G.~T.}\ \bibnamefont
			{Silva}},\ }\bibfield  {title} {\bibinfo {title} {{Generalization of the
				extended optical theorem for scalar arbitrary-shape acoustical beams in
				spherical coordinates.}},\ }\href@noop {} {\bibfield  {journal} {\bibinfo
			{journal} {Phys. Rev. E}\ }\textbf {\bibinfo {volume} {90}},\ \bibinfo
		{pages} {053204} (\bibinfo {year} {2014})}\BibitemShut {NoStop}%
	\bibitem [{\citenamefont {Silva}\ \emph {et~al.}(2015)\citenamefont {Silva},
		\citenamefont {Baggio}, \citenamefont {Lopes},\ and\ \citenamefont
		{Mitri}}]{Silva2015a}%
	\BibitemOpen
	\bibfield  {author} {\bibinfo {author} {\bibfnamefont {G.~T.}\ \bibnamefont
			{Silva}}, \bibinfo {author} {\bibfnamefont {A.~L.}\ \bibnamefont {Baggio}},
		\bibinfo {author} {\bibfnamefont {J.~H.}\ \bibnamefont {Lopes}},\ and\
		\bibinfo {author} {\bibfnamefont {F.~G.}\ \bibnamefont {Mitri}},\ }\bibfield
	{title} {\bibinfo {title} {Computing the acoustic radiation force exerted on
			a sphere using the translational addition theorem},\ }\href@noop {}
	{\bibfield  {journal} {\bibinfo  {journal} {IEEE Trans. Ultrason.
				Ferroelectr. Freq. Control}\ }\textbf {\bibinfo {volume} {62}},\ \bibinfo
		{pages} {576} (\bibinfo {year} {2015})}\BibitemShut {NoStop}%
	\bibitem [{\citenamefont {Gong}\ and\ \citenamefont
		{Marston}(2017)}]{Gong2017}%
	\BibitemOpen
	\bibfield  {author} {\bibinfo {author} {\bibfnamefont {Z.}~\bibnamefont
			{Gong}}\ and\ \bibinfo {author} {\bibfnamefont {P.~L.}\ \bibnamefont
			{Marston}},\ }\bibfield  {title} {\bibinfo {title} {Multipole expansion of
			acoustical {Bessel} beams with arbitrary order and location},\ }\href@noop {}
	{\bibfield  {journal} {\bibinfo  {journal} {J. Acoust. Soc. Am.}\ }\textbf
		{\bibinfo {volume} {141}},\ \bibinfo {pages} {EL574} (\bibinfo {year}
		{2017})}\BibitemShut {NoStop}%
	\bibitem [{\citenamefont {{Le\~ao}-Neto}\ \emph {et~al.}(2017)\citenamefont
		{{Le\~ao}-Neto}, \citenamefont {Lopes},\ and\ \citenamefont
		{Silva}}]{Leao-Neto2017}%
	\BibitemOpen
	\bibfield  {author} {\bibinfo {author} {\bibfnamefont {J.~P.}\ \bibnamefont
			{{Le\~ao}-Neto}}, \bibinfo {author} {\bibfnamefont {J.~H.}\ \bibnamefont
			{Lopes}},\ and\ \bibinfo {author} {\bibfnamefont {G.~T.}\ \bibnamefont
			{Silva}},\ }\bibfield  {title} {\bibinfo {title} {Extended optical theorem in
			isotropic solids and its application to the elastic radiation force},\
	}\href@noop {} {\bibfield  {journal} {\bibinfo  {journal} {J. Appl. Phys.}\
		}\textbf {\bibinfo {volume} {121}},\ \bibinfo {pages} {144902} (\bibinfo
		{year} {2017})}\BibitemShut {NoStop}%
	\bibitem [{\citenamefont {Zhang}(2018)}]{Zhang2018a}%
	\BibitemOpen
	\bibfield  {author} {\bibinfo {author} {\bibfnamefont {L.}~\bibnamefont
			{Zhang}},\ }\bibfield  {title} {\bibinfo {title} {A general theory of
			arbitrary {Bessel} beam scattering and interactions with a sphere},\
	}\href@noop {} {\bibfield  {journal} {\bibinfo  {journal} {J. Acoust. Soc.
				Am.}\ }\textbf {\bibinfo {volume} {143}},\ \bibinfo {pages} {2796} (\bibinfo
		{year} {2018})}\BibitemShut {NoStop}%
	\bibitem [{\citenamefont {Silva}(2011)}]{Silva2011a}%
	\BibitemOpen
	\bibfield  {author} {\bibinfo {author} {\bibfnamefont {G.~T.}\ \bibnamefont
			{Silva}},\ }\bibfield  {title} {\bibinfo {title} {{Off-axis scattering of an
				ultrasound Bessel beam by a sphere}},\ }\href@noop {} {\bibfield  {journal}
		{\bibinfo  {journal} {IEEE Trans. Ultrason. Ferroelectr. Freq. Control}\
		}\textbf {\bibinfo {volume} {58}},\ \bibinfo {pages} {298} (\bibinfo {year}
		{2011})}\BibitemShut {NoStop}%
	\bibitem [{\citenamefont {Mitri}\ and\ \citenamefont
		{Silva}(2011)}]{Mitri2011}%
	\BibitemOpen
	\bibfield  {author} {\bibinfo {author} {\bibfnamefont {F.~G.}\ \bibnamefont
			{Mitri}}\ and\ \bibinfo {author} {\bibfnamefont {G.~T.}\ \bibnamefont
			{Silva}},\ }\bibfield  {title} {\bibinfo {title} {{Off-axial acoustic
				scattering of a high-order Bessel vortex beam by a rigid sphere}},\
	}\href@noop {} {\bibfield  {journal} {\bibinfo  {journal} {Wave Motion}\
		}\textbf {\bibinfo {volume} {46}},\ \bibinfo {pages} {392} (\bibinfo {year}
		{2011})}\BibitemShut {NoStop}%
	\bibitem [{\citenamefont {Silva}\ \emph {et~al.}(2013)\citenamefont {Silva},
		\citenamefont {Lopes},\ and\ \citenamefont {Mitri}}]{Silva2013}%
	\BibitemOpen
	\bibfield  {author} {\bibinfo {author} {\bibfnamefont {G.~T.}\ \bibnamefont
			{Silva}}, \bibinfo {author} {\bibfnamefont {J.~H.}\ \bibnamefont {Lopes}},\
		and\ \bibinfo {author} {\bibfnamefont {F.~G.}\ \bibnamefont {Mitri}},\
	}\bibfield  {title} {\bibinfo {title} {{Off-axial acoustic radiation force of
				repulsor and tractor bessel beams on a sphere.}},\ }\href
	{https://doi.org/10.1109/TUFFC.2013.2683} {\bibfield  {journal} {\bibinfo
			{journal} {IEEE Trans. Ultrason. Ferroelectr. Freq. Control}\ }\textbf
		{\bibinfo {volume} {60}},\ \bibinfo {pages} {1207} (\bibinfo {year}
		{2013})}\BibitemShut {NoStop}%
	\bibitem [{\citenamefont {Lopes}\ \emph {et~al.}(2016)\citenamefont {Lopes},
		\citenamefont {Azarpeyvand},\ and\ \citenamefont {Silva}}]{Lopes2016}%
	\BibitemOpen
	\bibfield  {author} {\bibinfo {author} {\bibfnamefont {J.~H.}\ \bibnamefont
			{Lopes}}, \bibinfo {author} {\bibfnamefont {M.}~\bibnamefont {Azarpeyvand}},\
		and\ \bibinfo {author} {\bibfnamefont {G.~T.}\ \bibnamefont {Silva}},\
	}\bibfield  {title} {\bibinfo {title} {Acoustic interaction forces and
			torques acting on suspended spheres in an ideal fluid},\ }\href@noop {}
	{\bibfield  {journal} {\bibinfo  {journal} {IEEE Trans. Ultrason.
				Ferroelectr. Freq. Control}\ }\textbf {\bibinfo {volume} {63}},\ \bibinfo
		{pages} {186} (\bibinfo {year} {2016})}\BibitemShut {NoStop}%
	\bibitem [{\citenamefont {Gong}\ \emph
		{et~al.}(2019{\natexlab{b}})\citenamefont {Gong}, \citenamefont {Marston},\
		and\ \citenamefont {Li}}]{Gong2019}%
	\BibitemOpen
	\bibfield  {author} {\bibinfo {author} {\bibfnamefont {Z.}~\bibnamefont
			{Gong}}, \bibinfo {author} {\bibfnamefont {P.~L.}\ \bibnamefont {Marston}},\
		and\ \bibinfo {author} {\bibfnamefont {W.}~\bibnamefont {Li}},\ }\bibfield
	{title} {\bibinfo {title} {Reversals of acoustic radiation torque in bessel
			beams using theoretical and numerical implementations in three dimensions},\
	}\href@noop {} {\bibfield  {journal} {\bibinfo  {journal} {Phys. Rev. Appl.}\
		}\textbf {\bibinfo {volume} {11}},\ \bibinfo {pages} {064022} (\bibinfo
		{year} {2019}{\natexlab{b}})}\BibitemShut {NoStop}%
	\bibitem [{\citenamefont {Zhao}\ \emph {et~al.}(2019)\citenamefont {Zhao},
		\citenamefont {Thomas},\ and\ \citenamefont {Marchiano}}]{Zhao2019}%
	\BibitemOpen
	\bibfield  {author} {\bibinfo {author} {\bibfnamefont {D.}~\bibnamefont
			{Zhao}}, \bibinfo {author} {\bibfnamefont {J.-L.}\ \bibnamefont {Thomas}},\
		and\ \bibinfo {author} {\bibfnamefont {R.}~\bibnamefont {Marchiano}},\
	}\bibfield  {title} {\bibinfo {title} {Computation of the radiation force
			exerted by the acoustic tweezers using pressure field measurements},\
	}\href@noop {} {\bibfield  {journal} {\bibinfo  {journal} {J. Acoust. Soc.
				Am.}\ }\textbf {\bibinfo {volume} {146}},\ \bibinfo {pages} {1650} (\bibinfo
		{year} {2019})}\BibitemShut {NoStop}%
	\bibitem [{\citenamefont {{L. P. Gor'kov}}(1962)}]{Gorkov1962}%
	\BibitemOpen
	\bibfield  {author} {\bibinfo {author} {\bibnamefont {{L. P. Gor'kov}}},\
	}\bibfield  {title} {\bibinfo {title} {{On the forces acting on a small
				particle in an acoustic field in an ideal fluid}},\ }\href@noop {} {\bibfield
		{journal} {\bibinfo  {journal} {Sov. Phys.-Dokl.}\ }\textbf {\bibinfo
			{volume} {6}},\ \bibinfo {pages} {773} (\bibinfo {year} {1962})}\BibitemShut
	{NoStop}%
	\bibitem [{\citenamefont {Silva}(2014)}]{Silva2014}%
	\BibitemOpen
	\bibfield  {author} {\bibinfo {author} {\bibfnamefont {G.~T.}\ \bibnamefont
			{Silva}},\ }\bibfield  {title} {\bibinfo {title} {{Acoustic radiation force
				and torque on an absorbing compressible particle in an inviscid fluid}},\
	}\href {https://doi.org/10.1121/1.4895691} {\bibfield  {journal} {\bibinfo
			{journal} {J. Acoust. Soc. Am.}\ }\textbf {\bibinfo {volume} {136}},\
		\bibinfo {pages} {2405} (\bibinfo {year} {2014})}\BibitemShut {NoStop}%
	\bibitem [{\citenamefont {{Le\~ao-Neto}}\ and\ \citenamefont
		{Silva}(2016)}]{Leao-Neto2016}%
	\BibitemOpen
	\bibfield  {author} {\bibinfo {author} {\bibfnamefont {J.~P.}\ \bibnamefont
			{{Le\~ao-Neto}}}\ and\ \bibinfo {author} {\bibfnamefont {G.~T.}\ \bibnamefont
			{Silva}},\ }\bibfield  {title} {\bibinfo {title} {Acoustic radiation force
			and torque exerted on a small viscoelastic particle in an ideal fluid},\
	}\href@noop {} {\bibfield  {journal} {\bibinfo  {journal} {Ultrasonics}\
		}\textbf {\bibinfo {volume} {71}},\ \bibinfo {pages} {1} (\bibinfo {year}
		{2016})}\BibitemShut {NoStop}%
	\bibitem [{\citenamefont {Zhang}\ and\ \citenamefont
		{Marston}(2014)}]{Zhang2014}%
	\BibitemOpen
	\bibfield  {author} {\bibinfo {author} {\bibfnamefont {L.}~\bibnamefont
			{Zhang}}\ and\ \bibinfo {author} {\bibfnamefont {P.}~\bibnamefont
			{Marston}},\ }\bibfield  {title} {\bibinfo {title} {Acoustic radiation torque
			on small objects in viscous fluids and connection with viscous dissipation},\
	}\href@noop {} {\bibfield  {journal} {\bibinfo  {journal} {J. Acoust. Soc.
				Am.}\ }\textbf {\bibinfo {volume} {136}},\ \bibinfo {pages} {2917} (\bibinfo
		{year} {2014})}\BibitemShut {NoStop}%
	\bibitem [{\citenamefont {Pierce}(2019)}]{Pierce2019}%
	\BibitemOpen
	\bibfield  {author} {\bibinfo {author} {\bibfnamefont {A.~D.}\ \bibnamefont
			{Pierce}},\ }\href@noop {} {\emph {\bibinfo {title} {Acoustics: An
				Introduction to Its Physical Principles and Applications}}},\ \bibinfo
	{edition} {3rd}\ ed.\ (\bibinfo  {publisher} {Springer},\ \bibinfo {address}
	{Switzerland},\ \bibinfo {year} {2019})\ \bibinfo {note} {p. 489}\BibitemShut
	{NoStop}%
	\bibitem [{\citenamefont {Shi}\ \emph {et~al.}(2019)\citenamefont {Shi},
		\citenamefont {Zhao}, \citenamefont {Long}, \citenamefont {Yang},
		\citenamefont {Wang}, \citenamefont {Chen}, \citenamefont {Ren},\ and\
		\citenamefont {Zhang}}]{Shi2019}%
	\BibitemOpen
	\bibfield  {author} {\bibinfo {author} {\bibfnamefont {C.}~\bibnamefont
			{Shi}}, \bibinfo {author} {\bibfnamefont {R.}~\bibnamefont {Zhao}}, \bibinfo
		{author} {\bibfnamefont {Y.}~\bibnamefont {Long}}, \bibinfo {author}
		{\bibfnamefont {S.}~\bibnamefont {Yang}}, \bibinfo {author} {\bibfnamefont
			{Y.}~\bibnamefont {Wang}}, \bibinfo {author} {\bibfnamefont {H.}~\bibnamefont
			{Chen}}, \bibinfo {author} {\bibfnamefont {J.}~\bibnamefont {Ren}},\ and\
		\bibinfo {author} {\bibfnamefont {X.}~\bibnamefont {Zhang}},\ }\bibfield
	{title} {\bibinfo {title} {Observation of acoustic spin},\ }\href@noop {}
	{\bibfield  {journal} {\bibinfo  {journal} {Nat. Sci. Rev.}\ }\textbf
		{\bibinfo {volume} {6}},\ \bibinfo {pages} {707} (\bibinfo {year}
		{2019})}\BibitemShut {NoStop}%
	\bibitem [{\citenamefont {Bliokh}\ and\ \citenamefont
		{Nori}(2019)}]{Bliokh2019}%
	\BibitemOpen
	\bibfield  {author} {\bibinfo {author} {\bibfnamefont {K.~Y.}\ \bibnamefont
			{Bliokh}}\ and\ \bibinfo {author} {\bibfnamefont {F.}~\bibnamefont {Nori}},\
	}\bibfield  {title} {\bibinfo {title} {Spin and orbital angular momenta of
			acoustic beams},\ }\href@noop {} {\bibfield  {journal} {\bibinfo  {journal}
			{Phys. Rev. B}\ }\textbf {\bibinfo {volume} {99}},\ \bibinfo {pages} {174310}
		(\bibinfo {year} {2019})}\BibitemShut {NoStop}%
	\bibitem [{\citenamefont {Silva}\ \emph {et~al.}(2012)\citenamefont {Silva},
		\citenamefont {Lobo},\ and\ \citenamefont {Mitri}}]{Silva2012}%
	\BibitemOpen
	\bibfield  {author} {\bibinfo {author} {\bibfnamefont {G.~T.}\ \bibnamefont
			{Silva}}, \bibinfo {author} {\bibfnamefont {T.~P.}\ \bibnamefont {Lobo}},\
		and\ \bibinfo {author} {\bibfnamefont {F.~G.}\ \bibnamefont {Mitri}},\
	}\bibfield  {title} {\bibinfo {title} {Radiation torque produced by an
			arbitrary acoustic wave},\ }\href@noop {} {\bibfield  {journal} {\bibinfo
			{journal} {Europhys. Lett.}\ }\textbf {\bibinfo {volume} {97}},\ \bibinfo
		{pages} {54003} (\bibinfo {year} {2012})}\BibitemShut {NoStop}%
	\bibitem [{\citenamefont {Zhang}\ and\ \citenamefont
		{Marston}(2013)}]{Zhang2013}%
	\BibitemOpen
	\bibfield  {author} {\bibinfo {author} {\bibfnamefont {L.}~\bibnamefont
			{Zhang}}\ and\ \bibinfo {author} {\bibfnamefont {P.~L.}\ \bibnamefont
			{Marston}},\ }\bibfield  {title} {\bibinfo {title} {Optical theorem for
			acoustic non-diffracting beams and application to radiation force and
			torque},\ }\href@noop {} {\bibfield  {journal} {\bibinfo  {journal} {Biomed.
				Opt. Expr.}\ }\textbf {\bibinfo {volume} {4}},\ \bibinfo {pages} {1610}
		(\bibinfo {year} {2013})}\BibitemShut {NoStop}%
	\bibitem [{\citenamefont {Zhang}\ and\ \citenamefont
		{Marston}(2011)}]{Zhang2011}%
	\BibitemOpen
	\bibfield  {author} {\bibinfo {author} {\bibfnamefont {L.}~\bibnamefont
			{Zhang}}\ and\ \bibinfo {author} {\bibfnamefont {P.~L.}\ \bibnamefont
			{Marston}},\ }\bibfield  {title} {\bibinfo {title} {{Geometrical
				interpretation of negative radiation forces of acoustical Bessel beams on
				spheres.}},\ }\href {https://doi.org/10.1103/PhysRevE.84.035601} {\bibfield
		{journal} {\bibinfo  {journal} {Phys. Rev. E}\ }\textbf {\bibinfo {volume}
			{84}},\ \bibinfo {pages} {035601(R)} (\bibinfo {year} {2011})}\BibitemShut
	{NoStop}%
	\bibitem [{\citenamefont {Silbiger}(1963)}]{Silbiger1963}%
	\BibitemOpen
	\bibfield  {author} {\bibinfo {author} {\bibfnamefont {A.}~\bibnamefont
			{Silbiger}},\ }\bibfield  {title} {\bibinfo {title} {Scattering of sound by
			an elastic prolate spheroid},\ }\href@noop {} {\bibfield  {journal} {\bibinfo
			{journal} {J. Acoust. Soc. Am.}\ }\textbf {\bibinfo {volume} {35}},\
		\bibinfo {pages} {564} (\bibinfo {year} {1963})}\BibitemShut {NoStop}%
	\bibitem [{\citenamefont {Xu}\ \emph {et~al.}(2012)\citenamefont {Xu},
		\citenamefont {Qiu},\ and\ \citenamefont {Liu}}]{Xu2012}%
	\BibitemOpen
	\bibfield  {author} {\bibinfo {author} {\bibfnamefont {S.}~\bibnamefont
			{Xu}}, \bibinfo {author} {\bibfnamefont {C.}~\bibnamefont {Qiu}},\ and\
		\bibinfo {author} {\bibfnamefont {Z.}~\bibnamefont {Liu}},\ }\bibfield
	{title} {\bibinfo {title} {Transversally stable acoustic pulling force
			produced by two crossed plane waves},\ }\href@noop {} {\bibfield  {journal}
		{\bibinfo  {journal} {Europhys. Lett.}\ }\textbf {\bibinfo {volume} {99}},\
		\bibinfo {pages} {44003} (\bibinfo {year} {2012})}\BibitemShut {NoStop}%
	\bibitem [{\citenamefont {Dron}\ and\ \citenamefont {Aider}(2012)}]{Dron2012}%
	\BibitemOpen
	\bibfield  {author} {\bibinfo {author} {\bibfnamefont {O.}~\bibnamefont
			{Dron}}\ and\ \bibinfo {author} {\bibfnamefont {J.-L.}\ \bibnamefont
			{Aider}},\ }\bibfield  {title} {\bibinfo {title} {Acoustic energy measurement
			for a standing acoustic wave in a micro-channel},\ }\href@noop {} {\bibfield
		{journal} {\bibinfo  {journal} {Europhys. Lett.}\ }\textbf {\bibinfo {volume}
			{97}},\ \bibinfo {pages} {44011} (\bibinfo {year} {2012})}\BibitemShut
	{NoStop}%
	\bibitem [{\citenamefont {Lee}\ and\ \citenamefont {Wang}(1989)}]{Lee1989}%
	\BibitemOpen
	\bibfield  {author} {\bibinfo {author} {\bibfnamefont {C.~P.}\ \bibnamefont
			{Lee}}\ and\ \bibinfo {author} {\bibfnamefont {T.~G.}\ \bibnamefont {Wang}},\
	}\bibfield  {title} {\bibinfo {title} {Near-boundary streaming around a small
			sphere due to two orthogonal standing waves},\ }\href@noop {} {\bibfield
		{journal} {\bibinfo  {journal} {J. Acoust. Soc. Am.}\ }\textbf {\bibinfo
			{volume} {85}},\ \bibinfo {pages} {1081} (\bibinfo {year}
		{1989})}\BibitemShut {NoStop}%
	\bibitem [{\citenamefont {{Le\~ao-Neto}}\ \emph {et~al.}(2016)\citenamefont
		{{Le\~ao-Neto}}, \citenamefont {Lopes},\ and\ \citenamefont
		{Silva}}]{Leao-Neto2016a}%
	\BibitemOpen
	\bibfield  {author} {\bibinfo {author} {\bibfnamefont {J.~P.}\ \bibnamefont
			{{Le\~ao-Neto}}}, \bibinfo {author} {\bibfnamefont {J.~H.}\ \bibnamefont
			{Lopes}},\ and\ \bibinfo {author} {\bibfnamefont {G.~T.}\ \bibnamefont
			{Silva}},\ }\bibfield  {title} {\bibinfo {title} {Core-shell particles that
			are unresponsive to acoustic radiation force},\ }\href@noop {} {\bibfield
		{journal} {\bibinfo  {journal} {Phys. Rev. Applied}\ }\textbf {\bibinfo
			{volume} {6}},\ \bibinfo {pages} {024025} (\bibinfo {year}
		{2016})}\BibitemShut {NoStop}%
\end{thebibliography}
%

\end{document}